\documentclass[usenatbib]{mnras}
\usepackage{epsfig,lscape,color,graphicx}
\usepackage{longtable} 

\newcommand{\etal}{et~al.} 
\newcommand{\ionhy}{H{\sc ii} }

\newcommand{\kms}{$\mbox{km~s}^{-1}$ }
\newcommand{\kmsns}{$\mbox{km~s}^{-1}$}

\newcommand{\twofig}[2]        
{
   \begin{center}
     \begin{minipage}[t]{0.5\textwidth}
         \psfig{file=eps/#1.eps,height=0.95\textwidth}
     \end{minipage}
     \hfill
     \begin{minipage}[t]{0.5\textwidth}
         \psfig{file=eps/#2.eps,height=0.95\textwidth}
     \end{minipage}
   \end{center}
}

\begin{document}

\title[84-GHz methanol masers]{84-GHz methanol masers, their relationship to 36-GHz methanol masers and their molecular environments}
\author[S.\ L.\ Breen \etal]{S.\ L. Breen,$^{1}$\thanks{Email: Shari.Breen@sydney.edu.au} Y. Contreras,$^{2}$ J.\ R. Dawson,$^{3}$ S.\ P. Ellingsen,$^{4}$ M.\ A. Voronkov,$^{5}$ \newauthor T.\ P. McCarthy$^{4,5}$  \\
  \\
  $^1$ Sydney Institute for Astronomy (SIfA), School of Physics, University of Sydney, NSW 2006, Australia;\\
   $^2$ Leiden Observatory, Leiden University, P.O. Box 9513, NL-2300 RA Leiden, The Netherlands;\\
  $^3$ Department of Physics and Astronomy and MQ Research Centre in Astronomy, Astrophysics and Astrophotonics, Macquarie University, \\ NSW 2109;\\
$^4$ School of Natural Sciences, University of Tasmania, Private Bag 37, Hobart, Tasmania 7001, Australia;\\
  $^5$ CSIRO Astronomy and Space Science, Australia Telescope National Facility, PO Box 76, Epping, NSW 1710, Australia}
 \maketitle
  
 \begin{abstract}
We present observations of the 36- (4$_{-1}$ $\rightarrow$ 3$_0$E) and 84-GHz (5$_{-1}$ $\rightarrow$ 4$_0$E) class I methanol maser transitions towards a sample of 94 known class I sites. These observations resulted in 93 and 92 detections in the 84- and 36-GHz transitions. While the majority of the 36-GHz sources have been previously reported, many of the sites are observed in the 84-GHz transition for the first time. The near-simultaneous observations of the two transitions revealed strikingly similar spectral profiles and a mean and median 36- to 84-GHz integrated flux density ratio of 2.6 and 1.4. 

Alongside the 36- and 84-GHz observations, we included rare class II methanol masers at 37.7-, 38.3-, 38.5-, 86.6- and 86.9-GHz, a number of recombination lines, and thermal molecular transitions. We detect one new site of 86.6- and 86.9-GHz methanol masers, as well as six maser candidates in one or more of 37.7-, 38.3-, 38.5-, 86.6- and 86.9-GHz methanol maser transitions. 

We detect a relatively higher rate of HC$_3$N compared to that reported by MALT90 (once the respective detection limits were taken into account) who targeted dense dust clumps, suggesting that the class I methanol maser targets incorporate a relatively higher number of warm protostellar sources. We further find that there are similar relationships between the integrated flux density of both class I transitions with the integrated intensity of HC$_3$N, HNC, HCO$^+$, HNC, SiO and H$^{13}$CO$^+$. We suggest that this indicates that the integrated flux density of the 36- and 84-GHz transitions are closely linked to the available gas volume.

\end{abstract}


\section{Introduction}
Methanol masers are important tracers of star formation, in part due to the complexity of the methanol molecule that results in numerous maser transitions, each prevalent within a slightly different range of physical conditions \citep[e.g][]{Cragg05,McEwen14,Leurini16}. The many transitions of methanol masers are empirically divided into two classes of sources \citep[e.g.][]{Batrla87,Menten91b}. Class I methanol masers are collisionally excited and tend to occupy the region surrounding an outflow or an expanding \ionhy region \citep[e.g.][]{Voronkov14}, while class II methanol masers are radiatively pumped and so are generally much more closely associated with the young star formation region \citep[e.g.][]{CasMMB10}. Whereas some class I methanol masers have been detected towards low-mass stars \citep[e.g][]{Kalenskii2010}, class II methanol masers (at least at 6.7-GHz) are exclusively associated with high-mass star formation regions \citep[e.g.][]{Minier03,Xu08,Breen13}. 

In recent years, a definitive, unbiased search for class II methanol masers at 6.7-GHz has been made in the Southern hemisphere \citep{CasMMB10,GreenMMB10,CasMMB11,Green12,Breen15}, detecting 972 maser sites, each of which have been searched for accompanying 12.2-GHz emission, resulting in a detection rate of 45.3 per cent \citep{BreenMMB12a,BreenMMB12b,BreenMMB14,Breen16}. Other, rarer class II methanol maser observations are becoming more prevalent, and significant samples have now been targeted for the 19.9-, 23.1-, 37.7-, 38.3-, 38.5-, 85.5-, 86.6-, 86.9-, 107.0- and 156.6-GHz transitions \citep[e.g.][]{Val'tts99,Ellingsen03,Ellingsen04,Cragg04,Caswell00,Ellingsen11,Umemoto07}. These more rare transitions trace less commonly found physical conditions and their presence either indicates a short-lived evolutionary phase in the star formation process, or unusual star formation regions. In some cases the rarer transitions can also be weaker \citep[although high flux density sources also exist; e.g.][]{Ellingsen18} adding further complexity to their detection. Recently, studies of the transitions at 37.7-, 38.3- and 38.5-GHz have lead to the suggestion that their short-lived presence may indicate the end of the class II methanol maser phase in high-mass star formation phase \citep{Ellingsen11,Ellingsen13} and the first high-resolution observations of these transitions have been recently made \citep{Ellingsen18}. Other transitions like the 86.6-GHz 7$_{2}$ $\rightarrow$ 6$_{3}$ A$^{-}$ and the 86.9-GHz 7$_{2}$ $\rightarrow$ 6$_{3}$ A$^{+}$ transitions are especially rare, even taking into account the relatively small number of searches that have been conducted. To date, maser emission in these transitions have only been detected towards G345.01+1.79, W3(OH) and W51-IRS1 \citep{Cragg01,Sutton01,Minier02,Ellingsen03}.

Class I methanol masers studies have also generally been limited to targeted observations \citep[e.g.][]{Kurtz04,Ellingsen05,Cyg09,Chen11,Voronkov14,GM16,RG17}, with the exception of a recently completed survey of 5 square degrees of the Southern galaxy in the 44-GHz transition \citep{Jordan15,Jordan17}, a small region towards the Galactic center in the 36-GHz transition \citep{YZ13}, and a large-scale search for the rare 23.4-GHz transition \citep[only one detection made in a relatively shallow search across a 100$^{\circ}$ $\times$ 1$^{\circ}$ region of the Galactic plane;][]{Voronkov11}. The most extensively studied lines are those at 36-, 44- and 95-GHz and have resulted in hundreds of detections across the Galaxy \citep[e.g][]{Jordan17,Voronkov14,Chen11}. The 5$_{-1}$ - 4$_0$ E class I methanol maser transition at 84-GHz is particularly poorly characterised, having only been observed in a small number of searches since the transition was first detected towards DR 21(OH), NGC2264 and OMC-2 \citep{BM88,Menten91}. The most extensive search for this transition was conducted by \citet{Kalenskii01} who targeted 51 class I methanol masers, detecting narrow maser-like emission towards 14 of their targets and quasi-thermal emission towards a further 34. An additional search by \citet{RG18} targeted 38 sites of 44-GHz methanol maser emission with the Large Millimeter Telescope and resulted in a detection rate of 74 per cent despite a velocity resolution of $\sim$100~\kmsns. Further, targeted observations of this transition have included a limited number of sources, revealing detections at the locations of both high- and low-mass star formation regions \citep[e.g.][]{Salii02,Kalenskii06}. Maser emission from the 84-GHz transition (5$_{-1}$ - 4$_0$ E) is expected to be similar to the more widely studied 36-GHz (4$_{-1}$ - 3$_0$ E) transition given that they are consecutive transitions in the same transition series. The more commonly observed class I masers at 44- (7$_0$ $\rightarrow$ 6$_1$ A$^+$) and 95-GHz (8$_0$ $\rightarrow$ 7$_1$ A$^+$) are also consecutive transitions in the same  (J+1)$_0$ $\rightarrow$ J$_1$ A$^+$ transition series. The 95-GHz transition has a high detection rate towards 44-GHz sources, but is usually about a factor of three weaker \citep[e.g][]{Val'tts00,McCarthy18}. 

Here we present a series of spectral line observations conducted with the Mopra radio telescope towards 94 class I methanol maser targets \citep{Kurtz04,Voronkov14}. The primary goal of the observations was to detect new sources of the poorly studied 84-GHz class I methanol maser transition, with quasi-simultaneous observations of the 36-GHz class I methanol maser line also included to obtain meaningful line ratios to inform maser pumping models and to allow comparisons with observations of extragalactic class I sources \citep[e.g.][]{McCarthy17,Ellingsen17,McCarthyIP}. Alongside these main target lines we were able to include a number of other, rare class II methanol maser lines (at 37.7-, 38.3-, 38.5-, 86.6- and 86.9-GHz) as well as a number of recombination lines and thermal lines, tracing dense and shocked gas, and allowing us to make some comparisons between the detected methanol masers and their environments. Given that the Mopra beam is 1.3$\arcmin$ and at 36~GHz and 0.6$\arcmin$ at 84~GHz, on the scale of whole clumps rather than individual star formation regions, we present the results and discussion with this in mind.

\section{Observations and data reduction}

\subsection{Targets}
Our observations targeted the locations of 94 known sites of class~I methanol maser emission, comprising the full \citet{Voronkov14} sample of 71 southern 36- and 44-GHz class I methanol masers along with a further 23 sources from the \citet{Kurtz04} sample of 44-GHz class I methanol masers (we excluded 14 sources from their sample as their declinations were north of +20 degrees). 

\citet{Voronkov14} targeted their 36- and 44-GHz class~I methanol maser observation towards known class I methanol masers south of a declination of $-$35 degrees, combining detections reported by \citet{Slysh94}, \citet{Val'tts00} and \citet{Ellingsen05}, which themselves targeted the locations of \ionhy regions, 6.7-GHz methanol maser emission and 22-GHz water maser emission. \citet{Kurtz04} directed their 44-GHz class I methanol maser observations towards a diverse sample of 44 star formation regions, including both very young regions devoid of accompanying UC\ionhy regions as well as those slightly more evolved sources with developed UC\ionhy regions. Our sample of 94 targets therefore represents a range of sources, initially detected in a range of selection methods and as such should not be dominated by a particular class of object.  

Both \citet{Voronkov14} and \citet{Kurtz04} provide spot maps of each of their sources, revealing the extent of the class I methanol maser emission in each case. We have targeted the centre of the class I maser emission region as reported by \citet{Voronkov14} for 71 of our targets and observed the averaged right ascension and declinations of the maser spot positions reported in \citet{Kurtz04} for the remaining 23. Our full target list, together with appropriate references, is given in Table~\ref{tab:84_36}.   

\subsection{Observations}

We conducted targeted observations at both 7 and 3mm towards the 94 class I methanol maser sites, using the Mopra 22-m radio telescope between 2018 April 30 and 2018 May 6. At both frequencies, the Mopra spectrometer (MOPS) was configured to record two orthogonal linear polarisations across 16 sub-bands, each covering 138 MHz with 4096 channels. During the 7mm observations these sub bands were distributed in the frequency range of 33077 to 40805~MHz and during the 3mm observations between 84468 and 92127~MHz, allowing us to observe up to 16 lines simultaneously at both 7- and 3mm. This configuration resulted in a velocity coverage of $\sim$1100~\kms and a native velocity resolution of $\sim$0.34~\kms at 36.2-GHz and a velocity coverage of 480~\kms and a native velocity resolution of $\sim$0.14~\kmsns. The targeted lines are listed in Table~\ref{tab:lines}, and includes the typical 1-$\sigma$ noise limits as well as the final velocity resolution (taking any smoothing into account).

At both frequency set ups, the pointing was corrected by observing a nearby SiO maser approximately once per hour, resulting in pointing uncertainties of less than 10$\arcsec$.  Each target was observed in a series of position-switched observations, with reference observations made $-$15$\arcmin$ in declination from each target. At 7mm we observed a reference, source, source, reference pattern once and spent one minute at each position, resulting in a total on source integration time of two minutes. At 3mm we repeated the same observation pattern twice, and spent two minutes at each position, making our total on source integration time eight minutes. At 7mm the system temperature was measured soley by a continuously switched noise diode, but at 3mm we also made a paddle measurements every 15 - 20 mins to achieve a calibrated antenna temperature. We estimate that the flux density/antenna temperature measurements are accurate to 20 per cent (taking into account residual pointing errors, opacity variations and primary calibration errors).

At the frequency of the 36-GHz methanol maser line, the HPBW of Mopra is 1.3$\arcmin$ and at the frequency of the 84-GHz methanol maser line it as 0.6$\arcmin$, which are both large enough to accommodate the majority of expected class I methanol maser distributions \citep{Voronkov14}. 


\begin{table*}

 \caption{Target spectral lines split into methanol masers (top), thermal molecular lines (middle) and radio recombination lines (bottom). The observed line is followed by the adopted rest frequency (with uncertainties listed for the maser transitions in units of the least significant figure), the velocity resolution (post processing), the K to Jy conversion factor used (where appropriate), the typical 1-$\sigma$ noise level (in Jy where a conversion factor is given otherwise in K), rest frequency reference, and, finally notes indicating the class of maser or drawing attention to lines that had an incomplete coverage in our sample due to a suboptimal frequency set up (``incomplete") and are therefore not observed for every source.}
  \begin{tabular}{llllllll} \hline
 \multicolumn{1}{l}{\bf Spectral line} &\multicolumn{1}{c}{\bf Rest}  & \multicolumn{1}{c}{\bf V$_{res.}$} & \multicolumn{1}{c}{\bf K to Jy} &	\multicolumn{1}{c}{\bf typical} &\multicolumn{1}{c}{\bf Reference} & \multicolumn{1}{c}{\bf notes}\\
  &\multicolumn{1}{c}{\bf frequency}	& \multicolumn{1}{c}{\bf (\kmsns)} &	\multicolumn{1}{c}{\bf factor}& \multicolumn{1}{c}{\bf noise}\\
& 	& {\bf (MHz)} & &\multicolumn{1}{c}{\bf (Jy or K)} \\
  \hline

CH$_3$OH 4$_{-1}$ $\rightarrow$ 3$_0$ E  &  36169.290(14)	&	0.34 & 13.6&0.8	& \citet{XL97} & 	class I	\\
CH$_3$OH 7$_{-2}$ $\rightarrow$ 8$_{-1}$ E	&  37703.696(13)	& 0.32 & 14& 0.8 & \citet{XL97} & 	class II\\
CH$_3$OH 6$_2$ $\rightarrow$ 5$_3$ A$^-$ &	38293.292(14)	& 0.32& 14& 0.8 &	 \citet{XL97} & 	class II\\
CH$_3$OH 6$_2$ $\rightarrow$ 5$_3$ A$^+$ & 38452.652(14)	& 0.32 & 14&  0.8 &		 \citet{XL97} & class II\\
CH$_3$OH 5$_{-1}$ $\rightarrow$ 4$_0$ E & 84521.169(10) & 0.20 & 16 		& 0.8	& \citet{Muller04} & class I \\
CH$_3$OH 7$_2$ $\rightarrow$ 6$_3$ A$^-$ & 86615.600(5) & 0.20  & 16		&0.8	& \citet{Muller04} &  class II \\
CH$_3$OH 7$_2$ $\rightarrow$ 6$_3$ A$^+$ & 86902.949(5) & 0.20  & 16		& 0.8	& \citet{Muller04} & class II\\ \hline
H$^{13}$CN 			& 	86339.9214 & 0.20 & -- & 0.04& \citet{splat07}& incomplete\\
H$^{13}$CO$^{+}$ (1$-$0)&   86754.2884 & 0.20 & --& 0.04	& \citet{splat07}\\ 
SiO (2$-$0) v=0		& 	86846.96 & 0.20 & -- & 0.04 & \citet{splat07}	\\ 
HCN (1$-$0)			& 	88631.847 & 0.19 & --	& 0.04 & \citet{splat07}\\  
CH$_3$OH 15$_3$ $\rightarrow$ 14$_4$ A$^-$ 	& 88940.09	& 0.19 & --	& 0.04	& \citet{splat07}\\ 
HCO$^+$ (1$-$0)		& 	89188.5247	& 0.19 & -- & 0.04& \citet{splat07}	\\
CH$_3$OH 8$_{-4}$ $\rightarrow$ 9$_{-3}$ E 		& 89505.808 & 0.19 & --	& 0.04	& \citet{splat07}\\ 
HNC	(1$-$0)			& 90663.568	& 0.19 & --	& 0.04	& \citet{splat07}\\ 
HC$_3$N (10$-$9)	& 90979.023	& 0.19 & -- & 0.04& \citet{splat07}\\ 
CH$_3$CN 5(1)$-$4(1)& 91985.3141 & 0.19	& -- & 	0.05& \citet{splat07} \\ 
\hline
H72$\beta$	&		33821.51 & 0.51 & -- & 0.04 & \citet{Lilley68} \\
H57$\alpha$	&		34596.39 & 0.50 & -- & 0.04 & \citet{Lilley68}\\	
H69$\beta$	&		38360.28 & 0.48 & -- & 0.04 & \citet{Lilley68}\\
H55$\alpha$ &		38473.36 & 0.45 & -- & 0.04 & \citet{Lilley68}\\
H42$\alpha$			&85688.40 	& 0.20 & --	& 0.06	& \citet{Lilley68}&  incomplete\\ 
H41$\alpha$			& 92034.45 & 	0.19 & -- &	0.06 &\citet{Lilley68}	\\ 
\hline
\end{tabular}\label{tab:lines}
\end{table*}

\subsection{Data reduction}

The data were processed using the ATNF Spectral Analysis Package (ASAP) using standard techniques for position switched observations. Alignment of the velocity channels was carried out during processing and the adopted rest frequencies are given in Table~\ref{tab:lines}. For some of the lines, data were Hanning smoothed during the processing and the resultant velocity resolutions are also given in Table~\ref{tab:lines}. The maser data were converted from antenna temperature to Janskys following \citet{Urquhart10} at 7mm, who give conversion factors of 13.6 and 14 for the 36-GHz methanol transition and the three class II transitions near 38-GHz, respectively. At 3mm the conversion factor was calculated using the main beam efficiency presented by \citet{Ladd05} which implies an antenna temperature to Jansky conversion factor of 16. For each of the maser transitions the conversion factor used is listed in Table~\ref{tab:lines}. Molecular and recombination lines are presented in units of antenna temperature (T$_A^*$).

Typical 1-$\sigma$ noise limits for each of the lines are given in Table~\ref{tab:lines} in units of Jy for the masers and antenna temperature for all other lines. In the case where 36- and 84-GHz emission is not detected, source specific 3-$\sigma$ detection limits are given in Table~\ref{tab:84_36} and likewise for the thermal molecular and recombination lines listed in Table~\ref{tab:thermal}.

In the case of the molecular and recombination lines, we fitted Gaussian profiles to the emission. In the simple cases for lines without hyperfine structure, a single Gaussian component was used to determine the peak antenna temperature, peak velocity, line width and integrated intensity. For HCN we simultaneously fit the hyperfine components and have presented the peak antenna temperature, peak velocity, line width of the main line and the combined integrated intensity, including the hyperfine components. In the case of CH$_3$CN we have fit each of the detected hyperfine components simultaneously but presented each of them individually. Given our limited signal-to-noise ratio, we have detected a range in the number of components from one right through to five.








\section{Results}

Observations of seven different methanol maser transitions, 10 molecular lines and six recombination lines have resulted in a rich data set, with many detections of the target lines. Given the large number of lines, we describe the results in subgroups. For some sources which require additional information to that which can be described in tables and spectra there are comments given in Section~\ref{sect:individual}.

\subsection{36- and 84-GHz methanol sources}

Of the 94 class I methanol masers targeted (previously characterised at either or both of the 36- and 44-GHz transitions), we have detected 92 in the 36-GHz transition and 93 in the 84-GHz transition. The two sources we failed to detect at 36-GHz were Mol77 and G\,45.07+0.13 which were both reported by \citet{Kurtz04} as 44-GHz class I methanol masers (Mol77 is also the only source where no emission was detected in the 84-GHz methanol line). The first of these was detected with a peak flux density of 0.57~Jy and the second was detected at 1.08~Jy by \citet{Kurtz04} during their observations in 1999 and 2000, respectively. Given that 44-GHz masers are generally stronger than accompanying emission in the 36-GHz transition \citep{Voronkov14}, and that our 3-$\sigma$ 36-GHz detection limits are higher than the reported 44-GHz peak flux densities (1.9 and 1.8~Jy for the 36 and 84-GHz transitions in Mol77, and 2.9~Jy for the 36-GHz transition in G\,45.07+0.13), their non-detection is expected.

The properties of both the 36- and 84-GHz masers are given in Table~\ref{tab:84_36}, including 3-$\sigma$ detection limits where appropriate. References to previously detected 84-GHz sources are given in the final column, indicating that only six of the 93 detections have been reported in the literature previously. References to previously detected 36-GHz masers are not explicitly given in Table~\ref{tab:84_36} but those 71 targets taken from \citet{Voronkov14} (indicated by a `$^1$' following the source name) were characterised at both 36- and 44-GHz in that work. The remaining 23 targets are taken from a 44-GHz methanol maser catalogue \citep{Kurtz04} for which few sources have been followed up at 36-GHz previously.

For each of the 94 class I methanol maser targets, spectra of the detected 84-GHz sources have been overlaid with the detected 36-GHz emission in Fig.~\ref{fig:84_spect}. For completeness we have included the three spectra where we fail to detect any emission. These spectra highlight that the structure of the two transitions are remarkably similar in almost all cases.

\begin{figure*}
	\psfig{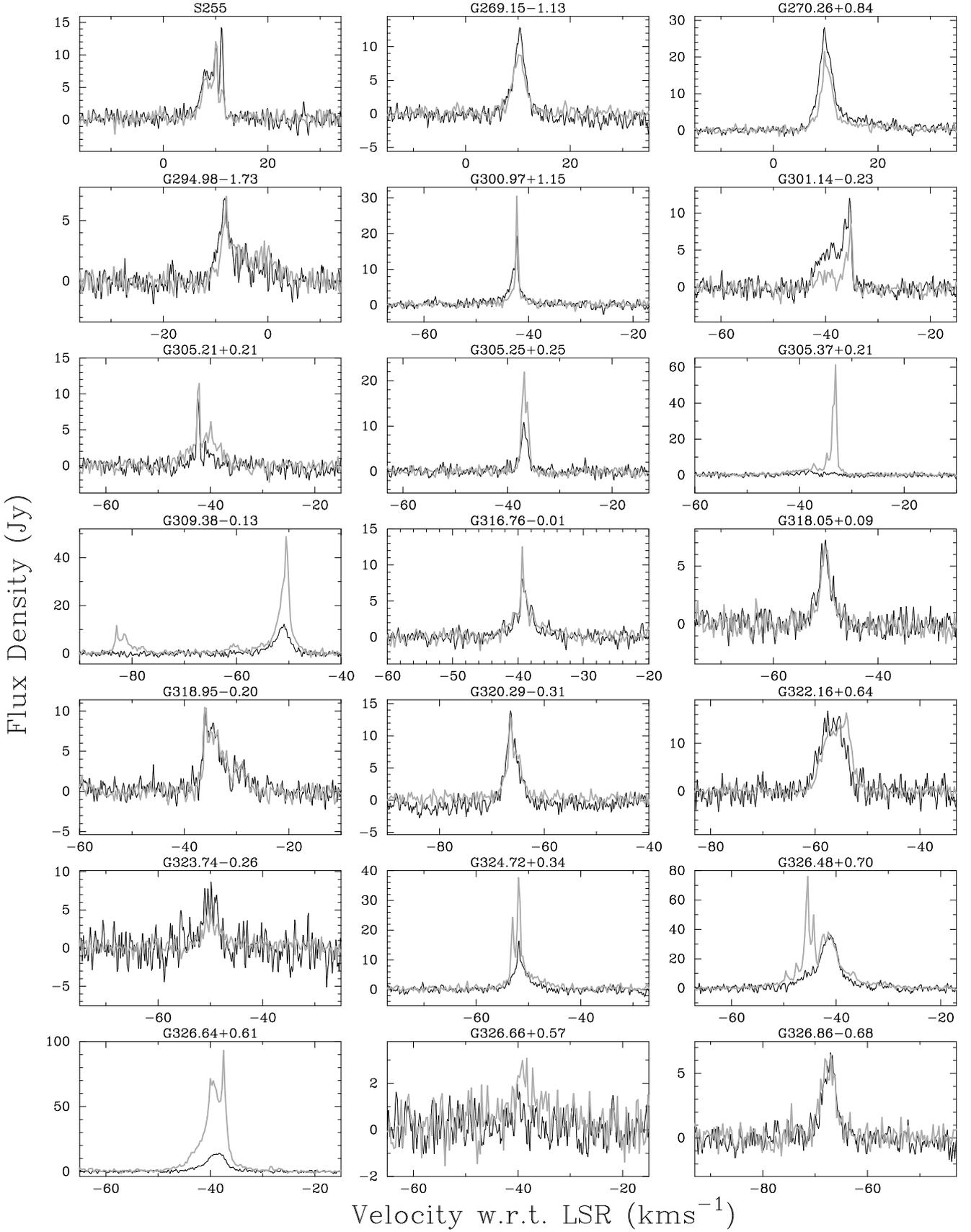}
\caption{Spectra of the 84-GHz methanol (black) and 36-GHz methanol (grey) sources detected towards class I methanol maser targets.}
\label{fig:84_spect}
\end{figure*}

\begin{figure*}\addtocounter{figure}{-1}
	\psfig{figure=84_meth_2overlay.eps}
	\caption{--{\emph {continued}}}
\end{figure*}

\begin{figure*}\addtocounter{figure}{-1}
	\psfig{figure=84_meth_3overlay.eps}
\caption{--{\emph {continued}}}
\end{figure*}

\begin{figure*}\addtocounter{figure}{-1}
	\psfig{figure=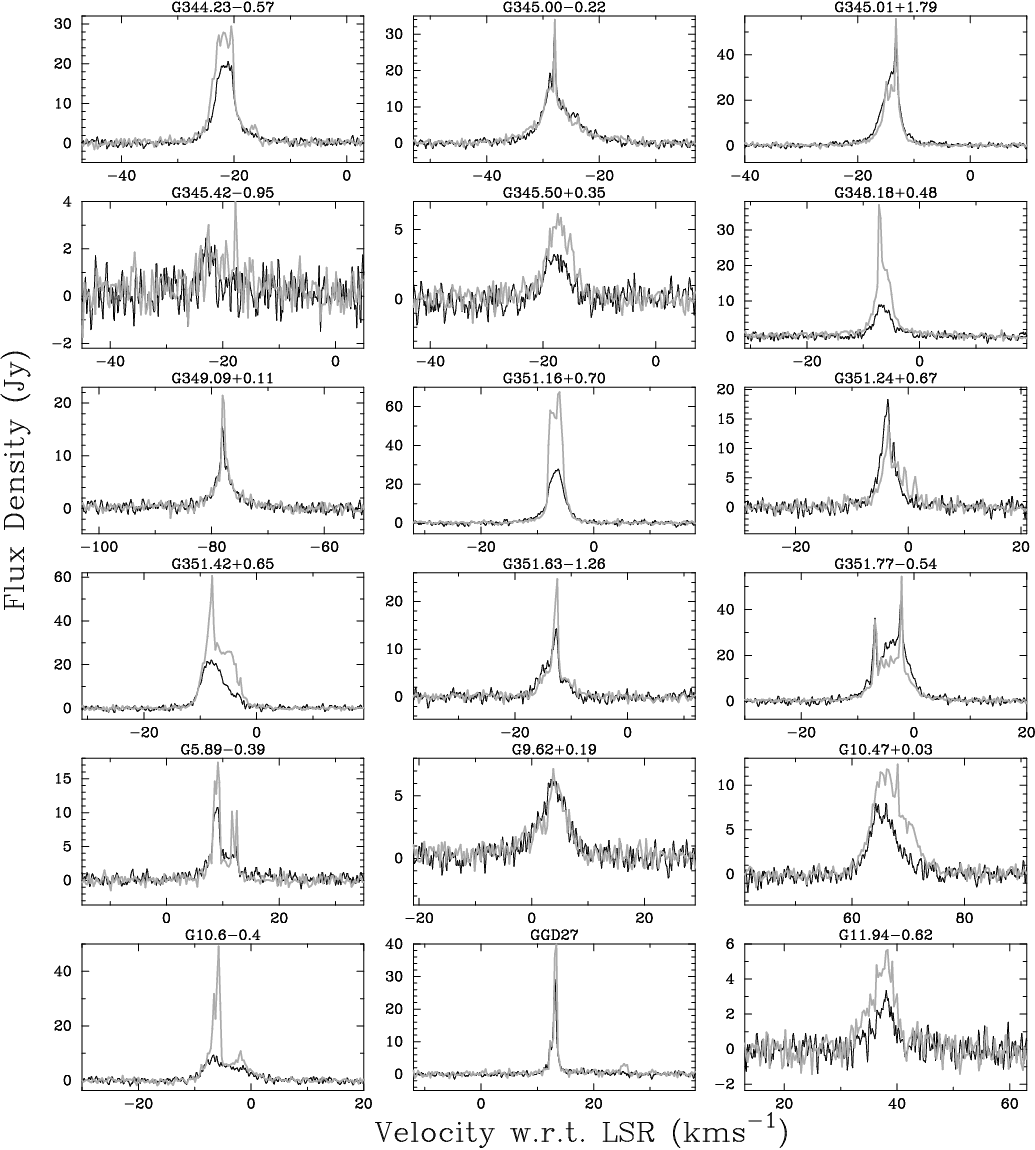}
\caption{--{\emph {continued}}}
\end{figure*}

\begin{figure*}\addtocounter{figure}{-1}
	\psfig{figure=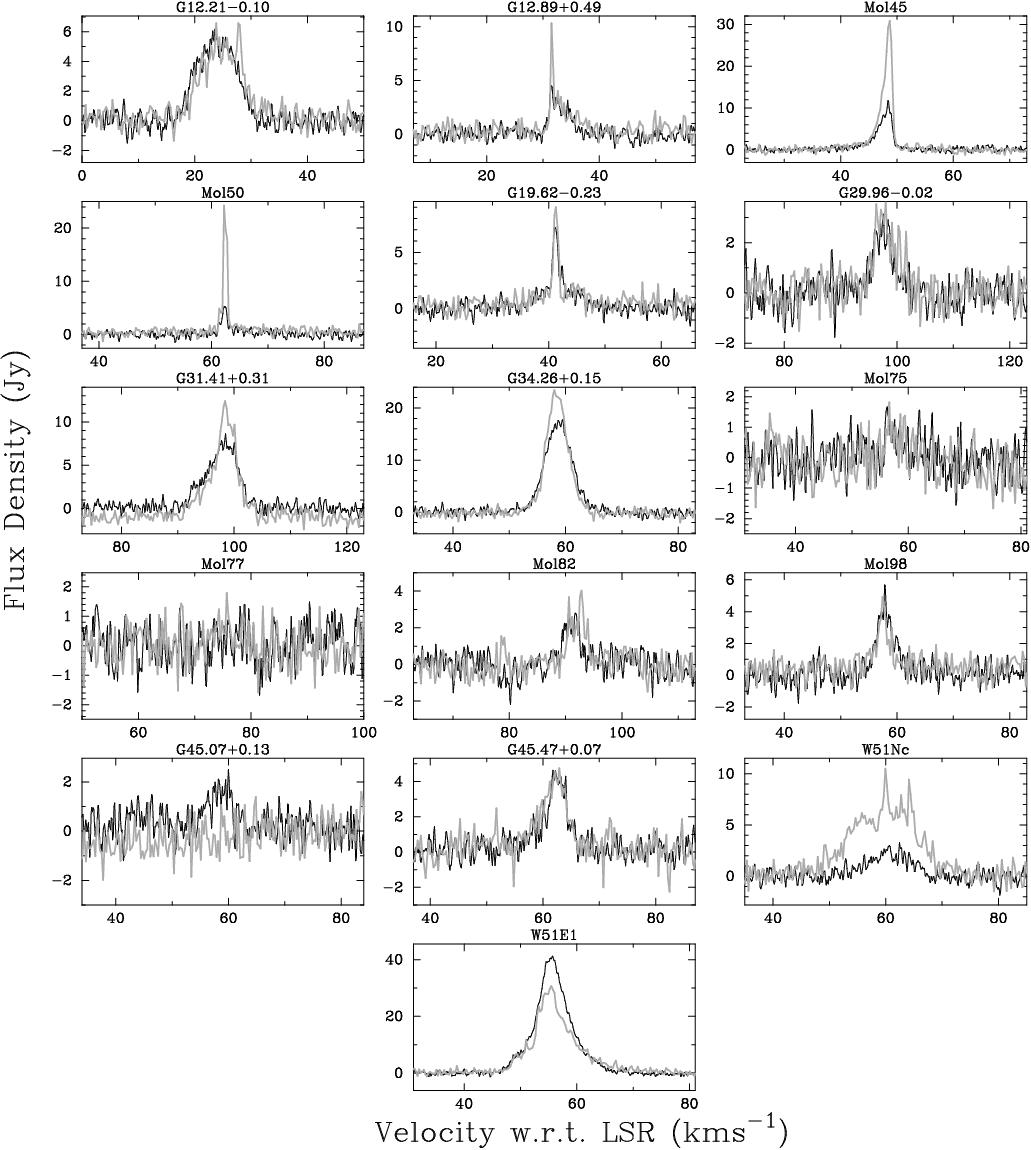}
\caption{--{\emph {continued}}}
\end{figure*}

\onecolumn

\begin{table*}
\caption{36- and 84-GHz methanol detections towards class I methanol maser sites reported in \citet{Voronkov14} and \citet{Kurtz04}. Source names (column 1) have been adopted from those publications, but where those names do not represent the Galactic coordinates, those are presented in parentheses following the adopted name. Column 2 and 3 give the targeted right ascension and declination (J2000), followed by two groups of five columns which give the minimum, maximum and peak velocity, the peak flux density and integrated flux density for the 36- and 84-GHz transitions (in units of Jy~\kmsns), respectively. In the case where transitions are not detected, 3-$\sigma$ detection limits are given in place of a measured peak flux density. 
Targets from the \citet{Voronkov14} sample were observed at 36-GHz in that work and references to previous detections of 84-GHz methanol emission are:  1: \citet{Kalenskii01}; 2: \citet{Voronkov06}.   
} 
\resizebox{\columnwidth}{!}{
  \begin{tabular}{llllllrlrlllcllllll} \hline
 \multicolumn{1}{l}{Source name} &\multicolumn{2}{c}{Equatorial coordinates} & \multicolumn{5}{c}{36-GHz methanol masers} & \multicolumn{5}{c}{84-GHz methanol masers}  & {Refs}\\
    \multicolumn{1}{l}{($l,b$)}& {RA (2000)} & {Dec. (2000)} & {V$_L$} & {V$_{H}$} & {V$_{pk}$} & {S$_{pk}$} & {S$_{int}$} & {V$_L$} & {V$_{H}$} & {V$_{pk}$} & {S$_{pk}$} &{S$_{int}$} \\	
      \multicolumn{1}{l}{($^{\circ}$ $^{\circ}$)}  &{(h m s)} & \multicolumn{1}{r}{($^{\circ}$ $'$ $''$)}&\multicolumn{3}{c}{(\kmsns)}& (Jy) &  &\multicolumn{3}{c}{(\kmsns)}& (Jy) & \\  \hline 
S255$^2$ (G\,192.58$-$0.04)	     	 &	06 12 53.8 &	+18 00 26.5	 & 6.4 & 11.7 & 10.0 & 12.0 & 27.7 & 6.1	&	11.8	& 11.1 & 14.2 & 37.9 	& 1 \\
G\,269.15$-$1.13$^1$ &	09 03 32.2 &	$-$48 28 10  & 8.1 & 12.3 & 10.1 & 8.8 & 24.2 & 7.6 & 13.0 & 10.3 & 12.8 & 31.0	\\
G\,270.26+0.84$^1$   &	09 16 40.7 &	$-$47 56 16  &	7.0 & 18.0 & 9.8 & 21.5 & 60.4 & 6.8 & 18.5 & 9.8 & 28.0 & 99.7\\
G\,294.98$-$1.73$^1$ &	11 39 14.1 &	$-$63 29 10  & $-$9.3 & 1.6 & $-$7.9 & 7.0 & 24.1 &	$-$10.6 & 2.2 & $-$8.2 & 6.8 & 21.9 \\
G\,300.97+1.15$^1$ &	12 34 52.3 &	$-$61 39 55  &	$-$43.6 & $-$40.6 & $-$42.2 & 30.5 & 21.4 & $-$44.5 & $-$40.7 & $-$42.2 & 19.3 & 26.7\\
G\,301.14$-$0.23$^1$ &	12 35 35.3 &	$-$63 02 29  &	$-$42.7 & $-$34.9 & $-$35.2 & 8.5 & 14.3 & $-$42.7 & $-$35.0 & $-$35.4 & 12.0 & 39.8 \\
G\,305.21+0.21$^1$ &	13 11 10.6 &	$-$62 34 45  & $-$45.8 & $-$36.8 & $-$42.1 &  11.4 & 28.1 & $-$43.5 & $-$39.4 & $-$42.4 & 9.3 & 9.5\\
G\,305.25+0.25$^1$ &	13 11 32.5 &	$-$62 32 08  &	$-$37.9 & $-$35.7 & $-$36.8 & 21.9 & 28.0 & $-$38.3 & $-$35.7 & $-$36.9 & 10.8 & 15.4\\
G\,305.37+0.21$^1$ &	13 12 33.1 &	$-$62 33 47  &	$-$43.1 & $-$31.1 & $-$33.1 & 61.2 & 84.8 & $-$39.8 & $-$37.5 & $-$38.1 & 2.7 & 3.7 \\
G\,309.38$-$0.13$^1$ &	13 47 22.3 &	$-$62 18 01  &	$-$83.5 & $-$46.9 & $-$50.5 & 48.7 & 142.5 &  $-$55.3 & $-$49.1 & $-$50.9 & 12.2 & 35.9\\
G\,316.76$-$0.01$^1$ &	14 44 56.1 &	$-$59 48 02  &	$-$42.4 & $-$35.7 & $-$39.4 & 12.5 & 19.9 & $-$42.4 & $-$35.2 & $-$39.3 & 8.1 & 16.2\\
G\,318.05+0.09$^1$ &	14 53 43.2 &	$-$59 08 58  & $-$52.4 & $-$47.4 & $-$50.2 & 6.9 & 14.7 &	$-$52.8 & $-$48.0 & $-$50.0 & 7.2 & 18.2\\
G\,318.95$-$0.20$^1$ &	15 00 55.5 &	$-$58 58 54  &	$-$37.2 & $-$27.7 & $-$36.1 & 10.4 & 40.6 & $-$36.9 & $-$27.8 & $-$36.1 & 10.0 & 38.7\\
G\,320.29$-$0.31$^1$ &	15 10 19.1 &	$-$58 25 17  &	$-$68.6 & $-$63.3 & $-$66.4 & 12.8 & 29.7 & $-$68.2 & $-$63.6 & $-$66.4 & 13.8 & 30.3\\
G\,322.16+0.64$^1$ &	15 18 37.0 &	$-$56 38 22  &	$-$61.0 & $-$50.7 & $-$54.0 & 16.3 & 85.3 & $-$61.0 & $-$50.6 & $-$57.6 & 16.8 & 85.8 \\
G\,323.74$-$0.26$^1$ &	15 31 45.6 &	$-$56 30 52  & $-$52.0 & $-$46.1 & $-$50.0 & 5.0 & 13.0 & 	$-$51.9 & $-$47.5 & $-$49.9 & 8.7 & 20.3\\
G\,324.72+0.34$^1$ &	15 34 57.9 &	$-$55 27 27  &	$-$55.5 & $-$45.2 & $-$51.9 & 37.6 & 73.9 & $-$54.3 & $-$47.1 & $-$51.8 & 16.4 & 33.8\\
G\,326.48+0.70$^1$ &	15 43 18.1 &	$-$54 07 25  & $-$54.9 & $-$30.1 & $-$45.4 & 76.0 & 337.7 &	$-$49.0 & $-$31.1 & $-$41.1 & 36.2 & 188.9\\
G\,326.64+0.61$^1$ &	15 44 31.0 &	$-$54 05 10  &	$-$48.7 & $-$26.3 & $-$37.5 & 93.1 & 368.5 & $-$44.3 & $-$35.3 & $-$38.3 & 13.9 & 64.6\\
G\,326.66+0.57$^1$ &	15 44 47.7 &	$-$54 06 43  & $-$41.1 & $-$37.2 & $-$38.3 &3.1 & 7.1 &$-$40.3 & $-$39.7 & $-$40.1 & 2.0 & 0.8 \\
G\,326.86$-$0.68$^1$ &	15 51 13.9 &	$-$54 58 05  &	$-$70.1 & $-$64.8 & $-$67.0 & 6.5 & 18.9 & $-$69.9 & $-$65.2 & $-67.1$ & 6.6 & 16.8\\
G\,327.29$-$0.58$^1$ &	15 53 09.5 &	$-$54 36 57  &	$-$51.6 & $-$37.7 & $-$45.5 & 61.8 & 263.4 & $-$51.6 & $-$38.4 & $-$45.6 & 68.4 & 304.8\\
G\,327.39+0.20$^1$ &	15 50 19.2 &	$-$53 57 07  &	$-$91.6 & $-$84.6 & $-$88.3 & 9.7 & 40.7 & $-$91.7 & $-$84.6 & $-$89.2 & 8.5 & 32.7\\
G\,327.62$-$0.11$^1$ &	15 52 50.2 &	$-$54 03 03  &	$-$91.7 & $-$85.8 & $-$88.4 & 3.9 & 10.3& $-$91.9 & $-$85.3 & $-$88.2 & 4.2 & 8.9\\
G\,328.21$-$0.59$^1$ &	15 57 59.5 &	$-$54 02 14  &	$-$42.3 & $-$40.4 & $-$40.7 & 11.2 & 10.4 & $-$42.1 & $-$39.7 & $-$40.9 & 8.5 & 9.1 \\
G\,328.24$-$0.55$^1$ &	15 58 01.2 &	$-$53 59 09  &	$-$47.4 & $-$37.9 & $-$40.9 & 6.3 & 27.4 & $-$46.5 & $-$41.5 & $-$42.0 & 2.5 & 2.0\\
G\,328.25$-$0.53$^1$ &	15 58 02.2 &	$-$53 57 29  &	$-$53.5 & $-$41.2 & $-$45.7 & 19.4 & 66.0 & $-$48.3 & $-$42.8 & $-$45.3 & 5.5 & 18.9\\
G\,328.81+0.63$^1$ &	15 55 48.7 &	$-$52 43 03  &	$-$46.1 & $-$38.5 & $-$42.4 & 17.7 & 48.6 & $-$56.7 & $-$21.9 & $-$41.4 & 22.6 & 104.1 \\
G\,329.03$-$0.20$^1$ &	16 00 30.9 &	$-$53 12 34  & $-$51.9 & $-$36.8 & $-$46.9 & 47.0 & 219.2 & $-$49.9 & $-$37.9 & $-$43.9 & 15.8 & 90.0\\
G\,329.07$-$0.31$^1$ &	16 01 09.6 &	$-$53 16 08  &	$-$49.6 & $-$34.8 & $-$44.0 & 28.0 & 162.7 & $-$48.5 & $-$38.0 & $-$44.0 & 10.3 & 55.9\\
G\,329.18$-$0.31$^1$ &	16 01 46.7 &	$-$53 11 38  &	$-$57.9 & $-$42.6 & $-$49.8 & 23.7 & 105.9 & $-$53.0 & $-$42.7 & $-$49.3 & 11.6 & 48.0\\
G\,329.47+0.50$^1$ &	15 59 39.8 &	$-$52 23 35  &	$-$79.9 & $-$60.7 & $-$69.1 & 11.2 & 71.4 & $-$75.4 & $-$66.1 & $-$68.6 & 5.0 & 22.6\\
G\,331.13$-$0.24$^1$ &	16 10 59.9 &	$-$51 50 19  &	$-$92.6 & $-$82.3 & $-$88.2 & 39.3 & 127.7 &$-$92.6 & $-$81.2 & $-$91.1 & 20.3 & 106.0\\
G\,331.34$-$0.35$^1$ &	16 12 26.5 &	$-$51 46 20  &	$-$66.2 & $-$65.1 & $-$65.4 & 15.2 & 11.4 & $-$66.6 & $-$65.2 & $-$65.6 & 8.1 & 7.7\\ 
G\,331.44$-$0.19$^1$ &	16 12 11.4 &	$-$51 35 09  &	$-$91.6 & $-$84.9 & $-$88.0 & 9.2 & 31.4 & $-$91.8 & $-$85.9 & $-$88.1 & 5.9 & 20.9\\
G\,332.30$-$0.09$^1$ &	16 15 45.2 &	$-$50 55 54  &	$-$53.3 & $-$47.5 &$-$49.7 & 14.7 & 21.8 & $-$53.6 & $-$46.2 & $-$49.8 & 10.8 & 30.1\\
G\,332.60$-$0.17$^1$ &	16 17 29.4 &	$-$50 46 08  &	$-$48.1 & $-$44.2 & $-$45.9 & 13.0 & 17.2 & $-$47.7 & $-$44.8 & $-$45.9 & 7.3 & 10.4\\
G\,332.94$-$0.69$^1$ &	16 21 20.3 &	$-$50 54 12  &	$-$52.2 & $-$47.2 & $-$49.5 & 3.3 & 6.2 & $-$50.8 & $-$47.4 & $-$48.8 & 4.2 & 7.9\\
G\,332.96$-$0.68$^1$ &	16 21 22.5 &	$-$50 52 57  &	$-$58.4 & $-$45.0 & $-$48.6 & 9.3 & 43.0 & $-$54.9 & $-$46.3 & $-$48.6 & 7.9 & 34.6\\
G\,333.03$-$0.06$^1$ &	16 18 56.7 &	$-$50 23 54  &	$-$40.9 & $-$39.7 & $-$40.9 & 2.9 & 2.6 & $-$42.0 & $-$39.8 & $-$40.9 & 2.5 & 3.0\\
G\,333.13$-$0.44$^1$ &	16 21 02.1 &	$-$50 35 49  &	$-$56.1 & $-$44.4 & $-$49.9 & 117.8 & 208.0 & $-$56.1 & $-$44.4 & $-$50.0 & 38.7 & 135.8\\
G\,333.13$-$0.56$^1$ &	16 21 36.1 &	$-$50 40 57  &	$-$63.9 & $-$49.4 & $-$56.1 & 37.7 & 171.0 & $-$62.9 & $-$51.7 & $-$56.0 & 16.1 & 83.1\\
G\,333.16$-$0.10$^1$ &	16 19 42.5 &	$-$50 19 57  &	$-$91.6 & $-$91.3 & $-$91.3 & 3.2 & 1.5 & $-$92.4 & $-$91.3 & $-$92.4 & 1.2 & 0.8 \\
G\,333.18$-$0.09$^1$ &	16 19 46.0 &	$-$50 18 34  &	$-$86.8 & $-$84.9 & $-$86.3 & 4.4 & 4.7 & $-$87.5 & $-$86.1 & $-$86.2 & 2.9 & 2.7\\
G\,333.23$-$0.06$^1$ &	16 19 50.8 &	$-$50 15 13  &	$-$98.0 & $-$79.0 & $-$87.1 & 95.9 & 248.0 & $-$91.9 & $-$83.3 & $-$87.2 & 48.0 & 120.1\\
G\,333.32+0.11$^1$ &	16 19 28.3 &	$-$50 04 46  &	$-$54.7 & $-$42.1 & $-$47.1 & 10.4 & 44.0 & $-$50.4 & $-$41.4 & $-$46.3 & 5.9 & 34.0\\
G\,333.47$-$0.16$^1$ &	16 21 20.2 &	$-$50 09 44  &	$-$47.8 & $-$40.6 & $-$43.1 & 21.9 & 43.9 & $-$47.4 & $-$42.0 & $-$43.0 & 4.3 & 9.0 \\
G\,333.56$-$0.02$^1$ &	16 21 08.8 &	$-$49 59 48  &	$-$40.3 & $-$38.9 & $-$39.7 & 28.7 & 22.4 & $-$40.3 & $-$39.0 & $-$40.0 & 5.4 & 4.8 \\
G\,333.59$-$0.21$^1$ &	16 22 06.7 &	$-$50 06 23  &	$-$51.9 & $-$44.4 & $-$49.4& 10.7 & 22.0 & $-$50.7 & $-$44.5 & $-$49.4 & 5.7 & 15.7\\
G\,335.06$-$0.43$^1$ &	16 29 23.4 &	$-$49 12 26  &	$-$41.2 & $-$37.0 & $-$40.6 & 7.5 & 17.8 & $-$43.3 & $-$38.3 & $-$39.5 & 5.8 & 9.3\\
G\,335.59$-$0.29$^1$ &	16 31 00.3 &	$-$48 43 37  &	$-$54.0 & $-$37.8 & $-45.3$ & 244.7 & 347.4 & $-$48.8 & $-$41.0 & $-$45.4 & 18.7 & 60.7\\
G\,335.79+0.17$^1$ &	16 29 46.5 &	$-$48 15 50  & $-$57.3 & $-$44.1 & $-$49.7 & 13.8 & 54.8 & $-$53.2 & $-$46.8 & $-$49.1 & 7.3 & 21.8	\\
G\,336.41$-$0.26$^1$ &	16 34 13.5 &	$-$48 06 18  &	$-$95.3 & $-$84.7 & $-$87.5 & 31.5 & 43.4 & $-$96.0 & $-$84.5 & $-$87.0 & 8.9 & 40.3\\
G\,337.40$-$0.40$^1$ &	16 38 48.9 &	$-$47 27 55  &	$-$45.8 & $-$36.8 & $-$40.5 & 11.9 & 22.5 & $-$43.3 & $-$38.9 & $-$41.8 & 8.1 & 18.1\\
G\,337.92$-$0.46$^1$ &	16 41 08.5 &	$-$47 07 44  &	$-$44.0 & $-$34.3 & $-$43.5 & 22.2 & 37.2 & $-$43.8 & $-$34.2 & $-$43.4 & 9.9 & 29.4 \\
G\,338.92+0.55$^1$ &	16 40 34.5 &	$-$45 41 50  &	$-$78.2 & $-$52.8 & $-$62.8 & 189.0 & 594.0 & $-$69.4 & $-$55.7 & $-$63.0 & 33.1 & 184.2\\
G\,339.88$-$1.26$^1$ &	16 52 04.3 &	$-$46 08 28  &	$-$34.5 & $-$30.1 & $-$31.5 & 6.1 & 14.6 & $-$34.1 & $-$31.1 & $-$31.8 & 3.7 & 5.2\\
G\,341.19$-$0.23$^1$ &	16 52 16.4 &	$-$44 28 44  &	$-$63.5 & $-$15.2 & $-$41.7 & 83.6 & 122.8 & $-$51.3 & $-$41.7 & $-$41.8 & 7.1 & 3.2\\

\end{tabular}%
}
\end{table*}

\begin{table*}\addtocounter{table}{-1}
  \caption{-- {\emph {continued}}} \label{tab:84_36}
 \resizebox{\columnwidth}{!}{
 \begin{tabular}{llllllrlrlllcllllll} \hline
 \multicolumn{1}{l}{Source name} &\multicolumn{2}{c}{Equatorial coordinates} & \multicolumn{5}{c}{36-GHz methanol masers} & \multicolumn{5}{c}{84-GHz methanol masers}  & {Refs}\\
    \multicolumn{1}{l}{($l,b$)}& {RA (2000)} & {Dec. (2000)} & {V$_L$} & {V$_{H}$} & {V$_{pk}$} & {S$_{pk}$} & {S$_{int}$} & {V$_L$} & {V$_{H}$} & {V$_{pk}$} & {S$_{pk}$} &{S$_{int}$} \\	
      \multicolumn{1}{l}{($^{\circ}$ $^{\circ}$)}  &{(h m s)} & \multicolumn{1}{r}{($^{\circ}$ $'$ $''$)}&\multicolumn{3}{c}{(\kmsns)}& (Jy) &  &\multicolumn{3}{c}{(\kmsns)}& (Jy) & \\  \hline 
G\,341.22$-$0.21$^1$ &	16 52 17.4 &	$-$44 27 03  &	$-$46.2 & $-$38.9 & $-$43.1 & 25.6 & 56.1 & $-$48.2 & $-$39.0 & $-$43.2 & 15.0 & 41.7\\
G\,343.12$-$0.06$^1$ &	16 58 16.8 &	$-$42 52 09  &	$-$36.6 & $-$22.9 & $-$27.1 & 43.3 & 115.7 & $-$36.5 & $-$24.1 & $-$27.3 & 23.4 & 79.6 & 2\\
G\,344.23$-$0.57$^1$ &	17 04 07.8 &	$-$42 18 26  &	$-$26.9 & $-$15.7 & $-$20.5 & 29.4 & 130.9 & $-$26.4 & $-$15.1 & $-$21.1 & 20.6 & 91.5\\
G\,345.00$-$0.22$^1$ &	17 05 10.7 &	$-$41 29 13  &	$-$34.9 & $-$20.1 & $-$27.9 & 34.0 & 100.3 & $-$34.4 & $-$20.1 & $-$28.0 & 29.6 & 102.0\\
G\,345.01+1.79$^1$ &	16 56 46.0 &	$-$40 14 09  &	$-$19.6 & $-$9.9 & $-$13.2 & 55.7 & 111.6 & $-$20.1 & $-$9.1 & $-$13.2 & 46.8 & 135.5\\
G\,345.42$-$0.95$^1$ &	17 09 35.6 &	$-$41 35 40  &	$-$24.8 & $-$17.5 & $-$17.8 & 4.0 & 8.9 & $-$23.8 & $-$21.3 & $-$23.0 & 2.4 & 3.4\\
G\,345.50+0.35$^1$ &	17 04 24.1 &	$-$40 44 07  &	$-$19.9 & $-$13.5 & $-$17.4 & 6.1 & 25.7 & $-$19.8 & $-$14.0 & $-$17.4 & 3.2 & 12.0\\
G\,348.18+0.48$^1$ &	17 12 06.8 &	$-$38 30 38  &	$-$10.0 & $-$1.3 & $-$7.2 & 37.1 & 82.1 & $-$9.5 & $-$2.8 & $-$6.8 & 8.9 & 30.8\\
G\,349.09+0.11$^1$ &	17 16 24.6 &	$-$37 59 43  &	$-$82.0 & $-$73.0 & $-$78.1 & 21.4 & 41.9 & $-$81.9 & $-$73.3 & $-$78.1 & 15.4 & 36.3\\
G\,351.16+0.70$^1$ &	17 19 56.4 &	$-$35 57 54  &	$-$11.4 & $-$1.6 & $-$6.1 & 67.4 & 185.4 & $-$11.8 & $-$2.3 & $-$6.3 & 27.8 & 101.3\\
G\,351.24+0.67$^1$ &	17 20 18.0 &	$-$35 54 45  &	$-$7.1 & 3.2 & $-$3.5 & 13.9 & 46.0 & $-$7.0 & 1.4 & $-$3.6 & 18.3 & 48.3\\
G\,351.42+0.65$^1$ &	17 20 52.3 &	$-$35 46 42  &	$-$11.8 & $-$1.2 & $-$7.9 & 60.6 & 210.9 & $-$11.8 & $-$1.5 & $-$8.1 & 22.0 & 111.1\\
G\,351.63$-$1.26$^1$ &	17 29 17.9 &	$-$36 40 29  &	$-$16.1 & $-$8.6 & $-$12.5 & 24.7 & 44.5 & $-$17.4 & $-$9.9 & $-$12.6 &14.2 & 39.5\\
G\,351.77$-$0.54$^1$ &	17 26 42.4 &	$-$36 09 14  &	$-$11.4 & 4.0 & $-$2.2 & 54.3 & 160.5 & $-$11.3 & 3.6 & $-$2.4 &43.5 & 206.1\\
G\,5.89$-$0.39$^2$   &	18 00 31.0 &	$-$24 03 52.4&	7.2 & 12.7 & 9.1 & 17.4 & 36.1 & 6.7 & 12.8 & 9.1 & 10.8 & 29.4 \\
G\,9.62+0.19$^2$   &	18 06 15.1 &	$-$20 31 37.1&	2.2 & 7.5 & 3.8 & 7.1 & 22.5 & 2.1 & 7.4 & 3.4 & 6.3 & 23.5\\
G\,10.47+0.03$^2$  &	18 08 37.9 &	$-$19 51 34.2&	61.1 & 74.6 & 68.1 & 12.3 & 79.8 & 60.2 & 73.0 & 66.0 & 7.9 & 39.9\\
G\,10.6$-$0.4$^2$    &	18 10 29.0 &	$-$19 55 46.4&	$-$9.1 & 2.1 & $-$5.7 & 49.1 & 106.5 & $-$9.3 & 2.2 & $-$6.6 & 9.3 & 51.1\\
GGD27$^2$ (G\,10.84$-$2.59)	     &	18 19 12.4 &	$-$20 47 24.8&	11.1 & 25.9 & 13.4 & 39.8 & 39.2 & 11.2 & 14.5 & 13.1 & 29.1 & 27.7 & \\
G\,11.94$-$0.62$^2$  &	18 14 02.2 &	$-$18 53 32.0&	32.7 & 40.6 & 38.3 & 5.7 & 23.4 & 35.1 & 39.4 & 38.1 & 3.4 & 7.6\\
G\,12.21$-$0.10$^2$  &	18 12 40.1 &	$-$18 24 21.4&	18.5 & 30.5 & 23.8 & 6.6 & 43.8 & 18.2 & 29.3 & 23.7 & 6.2 & 43.8\\
G\,12.89+0.49$^2$  &	18 11 50.8 &	$-$17 31 35.9&	31.2 & 36.3 & 31.5 & 10.3 & 13.3 & 31.0 & 35.6 & 31. 5 & 4.5 & 10.5 \\
Mol45$^2$ (G\,13.66$-$0.60)	     &	18 17 23.6 &	$-$17 22 13.0&	42.6 & 53.7 & 48.7 & 30.8 & 64.1 & 43.7 & 49.6 & 48.4 & 11.7 & 28.1 & \\
Mol50$^2$ (G\,14.89$-$0.40)	     &	18 19 07.6 &	$-$16 11 25.6&	61.4 & 64.4 & 62.2 & 24.2 & 23.8 & 61.2 & 64.4 & 62.2 & 5.3 & 6.7 &  \\
G\,19.62$-$0.23$^2$  &	18 27 37.7 &	$-$11 56 36.5&	36.5	&	46.3&41.3	& 9.0	& 16.0	&	37.8&45.9	&41.1 & 7.2 & 15.9	& 1\\
G\,29.96$-$0.02$^2$  &	18 46 03.1 &	$-$02 39 26.2&	95.2	& 101.6	&98.0	& 3.6	& 11.8	& 95.6	& 99.5	&	97.5& 3.1&7.5 & 1\\
G\,31.41+0.31$^2$  &	18 47 34.4 &	$-$01 12 48.8&	93.4	& 101.5	&98.4 	& 12.4	& 47.6	& 92.3	& 102.1	&98.5 & 8.6 & 47.9	& 1\\
G\,34.26+0.15$^2$  &	18 53 17.4 &	+01 15 04.6&	53.8	& 63.9	&58.0	& 23.5	& 106.3 	& 53.4	& 64.0	& 59.5 & 17.8 & 99.3	& 1\\
Mol75$^2$ (G\,34.82+0.35)	     &	18 53 37.7 &	+01 50 25.4& 56.6 & 	56.8 & 56.6 & 1.8 & 0.5 & 55.9 & 56.6 & 56.3 & 1.7 & 1.2 & \\
Mol77$^2$ (G\,36.12+0.55)	     &	18 55 16.8 &	+03 05 06.7& & & & $<$1.9 & & & & & $<$1.8 & & 	\\
Mol82$^2$ (G\,37.27+0.08)	     &	18 59 03.7 &	+03 53 42.9&	78.6 & 93.9 & 92.8 & 4.0 & 8.5 & 89.4 & 92.4 & 90.6 & 3.1 & 6.2 & \\
Mol98$^2$ (G\,43.04$-$0.45)	     &	19 11 38.9 &	+08 46 34.0&	56.0 & 59.1 & 57.4 & 4.9 & 8.7 & 55.6 & 60.5 & 57.8 & 5.7 & 14.4 & \\
G\,45.07+0.13$^2$  &	19 13 22.0 &	+10 50 59.0& & & & $<$2.9 & & 56.4 & 60.6 & 60.0 & 2.5 & 4.2	\\
G\,45.47+0.07$^2$  &	19 14 25.8 &	+11 09 27.4&	57.9 & 65.1 & 62.9 & 4.7 & 20.7 & 58.3 & 65.6 & 61.8 & 4.6 & 16.2\\
W51E1$^2$ (G\,49.49$-$0.39)	     &	19 23 44.3 &	+14 30 36.7& 48.2 & 65.5 & 55.5 & 30.7 & 219.7 & 47.0 & 64.9 & 55.7 & 41.2 & 285.1	&  1\\ 
W51Nc$^2$ (G\,49.49$-$0.37)	     &	19 23 40.1 &	+14 31 13.8&	47.9	& 71.7 	& 59.9	& 10.5	& 95.4	& 52.8	& 66.0	& 62.4 & 3.3 & 18.3	& \\
\hline

\end{tabular}%
}
\end{table*}

\twocolumn

\subsection{37.7-, 38.3- and 38.5-GHz methanol masers}

\begin{figure*}
	\epsfig{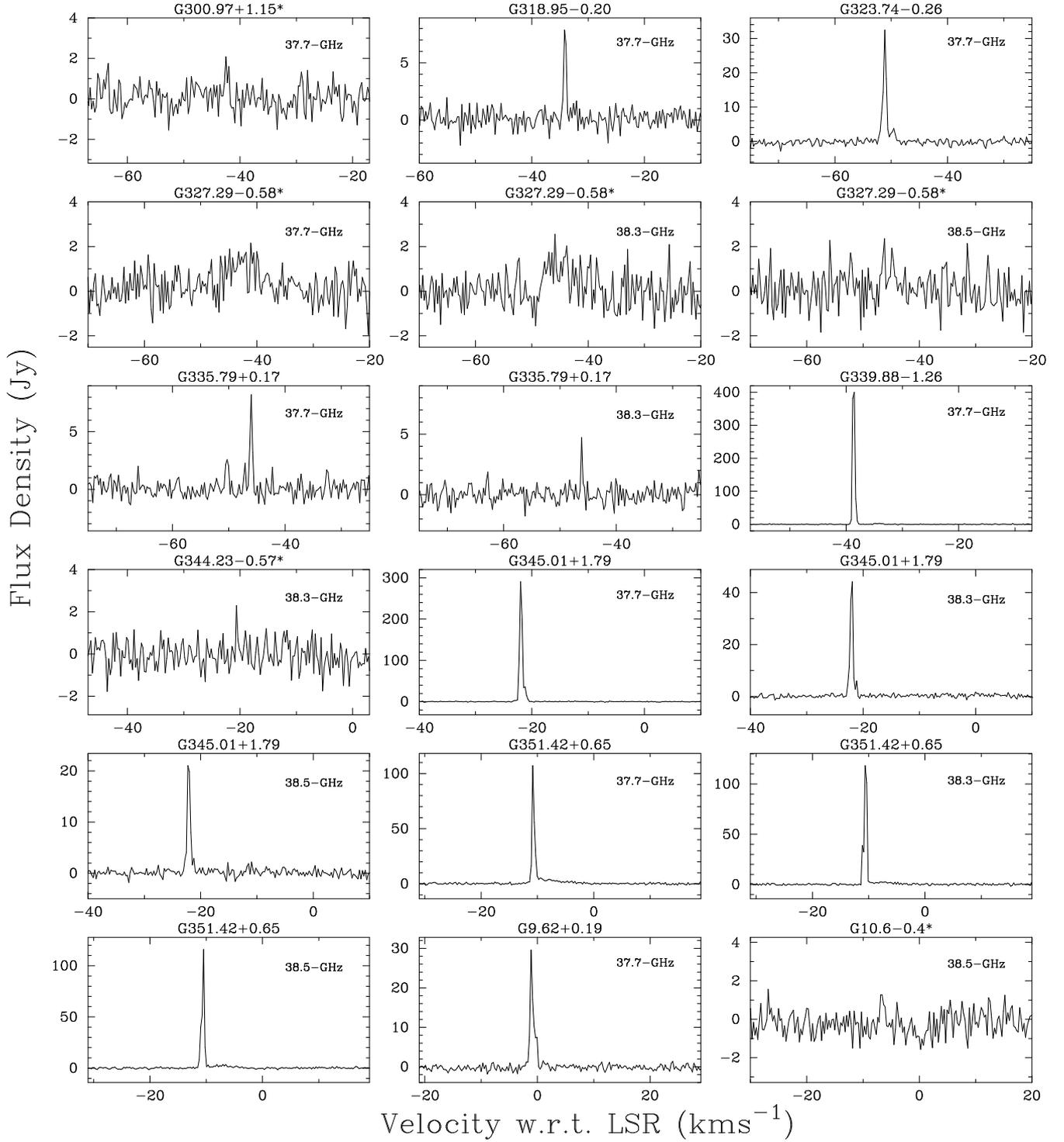}
\caption{Spectra of the 37.7-, 38.3- and 38.5-GHz sources detected towards class I methanol maser sources. New candidate detections are marked with a `*' following the source name.}
\label{fig:37_spect}
\end{figure*}

Alongside our observations of class I 36-GHz methanol masers we were able to simultaneously observe the rarer, 37.7-, 38.3- and 38.5-GHz class II methanol maser lines. Despite the 7~mm observations being motivated by the (generally) much stronger 36-GHz transition which required very short on source integration times (2mins), we were able to detect emission from a number of maser lines. In total we detect seven known 37.7-GHz masers, three known 38.3-GHz masers and two known 38.5-GHz masers \citep{Ellingsen11,Ellingsen13,Ellingsen18,Haschick89}. While these known masers account for most of our detections, we also present a further six maser candidates that are very close to the 3-$\sigma$ detection limit of the observations. While their peak flux densities are low, they all have velocities close to both the detected 36-GHz class I methanol maser emission and the velocity ranges of reported 6.7-GHz methanol maser emission \citep{GreenMMB10,CasMMB11,Green12}. The properties of the previously detected sources, together with our six maser candidates, are presented in Table~\ref{tab:7mm_meth}. Spectra for each of the listed sources are given in Fig.~\ref{fig:37_spect}. 

Of the six maser candidates, three are accounted for by G\,327.29$-$0.58 which shows very weak potential emission in each of the three transitions. If these transitions were considered in isolation the `emission' would not be notable, but given the marginal detections at the same velocity in each of the transitions it is worth including in more sensitive follow-up observations. Further discussion of this source, together with the other marginal detections, is presented in Section~\ref{sect:individual}.

\onecolumn
\begin{table}

\caption{Characteristics of methanol masers detected at 37.7-, 38.3- and 38.5-GHz. The first column gives the class I methanol maser source name and the next three groups of five columns give the minimum, maximum and peak velocity, the peak flux density and integrated flux density for the 37.7-, 38.3- and 38.5-GHz transitions, respectively. In the case where transitions are not detected, 3-$\sigma$ detection limits are given. All new maser candidates are close to the 3-$\sigma$ detection limit and have been marked with an `*' following the peak flux density. These are discussed individually in Section~\ref{sect:individual}. References to previously detected sources are: 1: \citet{Ellingsen11}; 2: \citet{Ellingsen13}; 3: \citet{Ellingsen18}; 4: \citet{Haschick89}.} \label{tab:7mm_meth}
\resizebox{\columnwidth}{!}{
\begin{tabular}{lllllllllllllllllllllllllllll} \hline
 \multicolumn{1}{l}{Source name} &\multicolumn{5}{c}{37.7GHz methanol} & \multicolumn{5}{c}{38.3-GHz methanol masers} & \multicolumn{5}{c}{38.5-GHz methanol masers} & Refs\\
    {($l,b$)}& {V$_L$} & {V$_{H}$} & {V$_{pk}$} & {S$_{pk}$} & {S$_{I}$} & {V$_L$} & {V$_{H}$} & {V$_{pk}$} & {S$_{pk}$} & {S$_{I}$} & {V$_L$} & {V$_{H}$} & {V$_{pk}$} & {S$_{pk}$} & {S$_{I}$}\\	
    {($^{\circ}$ $^{\circ}$)} & \multicolumn{3}{c}{(\kmsns)} & (Jy) &  & \multicolumn{3}{c}{(\kmsns)}& (Jy) &  & \multicolumn{3}{c}{(\kmsns)}& (Jy) & \\  \hline 
G\,300.97+1.15 		&   $-$42.5 & $-$42.0 &	$-$42.5 & 2.1* & 1.0 & & & & $<$2.1 & & & & & $<$2.1 \\
G\,318.95$-$0.20 	&	$-$34.5	& $-$33.9 & $-$34.2 & 7.9 & 4.4	& & & & $<$2.3 & & & & & $<$2.3 & & 1,2,3\\
G\,323.74$-$0.26 	&	$-$52.2 & $-$49.3 & $-$51.1 & 32.4 & 24.9 & & & & $<$2.3 & & & & & $<$2.3 & & 1,2,3\\
G\,327.29$-$0.58    &	$-$46.5 & $-$40.0 & $-$41.1 & 2.2* & 4.6 & $-$47.0 & $-$43.8 & $-$45.9 & 2.6* & 2.9 & $-$46.5 & $-$44.9 & $-$46.2 & 2.3* & 1.5 \\
G\,335.79+0.17		&	$-$50.6 & $-$45.7 & $-$46.0 & 8.2  & 6.5	& $-$46.6 & $-$46.1 & $-$46.4 & 4.7 & 1.6 &&&&$<$2.0&& 2\\
G\,339.88$-$1.26 &		$-$39.4 & $-$33.5 & $-$38.6 & 400 & 242	&&&&$<$1.9&&&&&$<$2.0&& 1,2,3\\
G\,344.23-0.57 & & & & $<$1.9& & $-$20.7 & $-$20.5 & $-$20.6 & 2.3* & 0.6 & & & & $<$2.0 \\
G\,345.01+1.79 		&	$-$23.3 & $-$20.7 & $-$22.0 & 291 & 186	& $-$22.9 & $-$21.2 & $-$22.1 & 44 & 30 & $-$22.8 & $-$21.2 & $-$22.3& 21.1&15.8& 1,2,3\\
G\,351.42+0.65 		&	$-$11.1 & $-$6.0  & $-$10.8	& 107 & 70 & $-$12.6 & $-$9.3 & $-$10.8 & 118 & 80  & $-$11.2& $-$6.3& $-$10.5 & 115.9 & 70.2& 1,2,3,4\\
G\,9.62+0.19  		&	$-$1.7	& 0.2	  & $-$1.1	& 30  & 22 &&&&$<$2.1&&&&&$<$2.1&& 1,2,3,4\\
G\,10.6$-$0.4	& & & & $<$2.0 & & & & & $<$2.0 & & $-$6.9 & $-$6.6 & $-$6.6 & 1.3* & 0.7\\ \hline 
\end{tabular}%
}
\end{table}

\twocolumn

\subsection{86.6- and 86.9-GHz methanol detections}

\begin{figure*}
	\epsfig{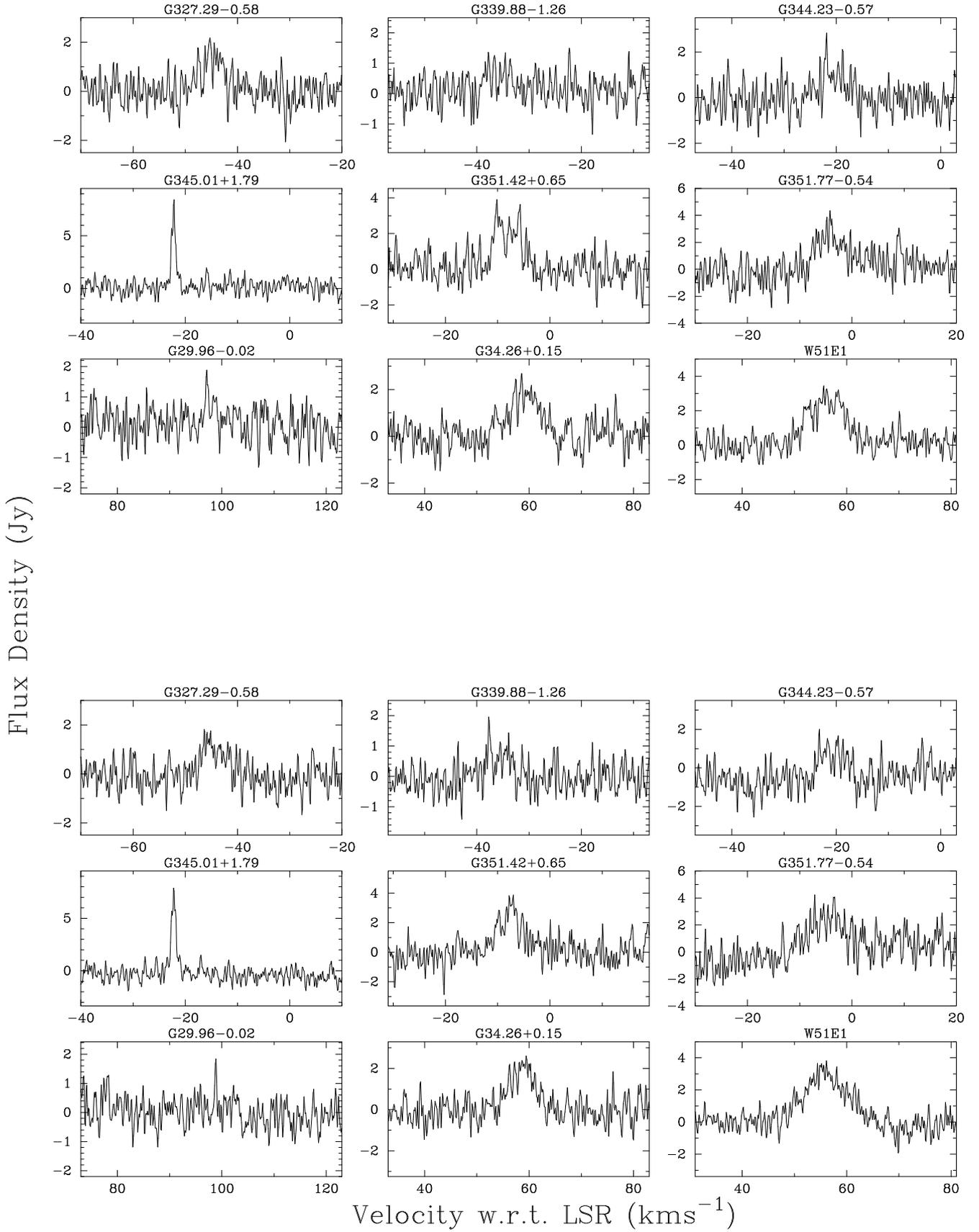}
\caption{Spectra of the 86.6-GHz (top) and 86.9-GHz (bottom) sources detected towards class I methanol masers.}
\label{fig:86_spect}
\end{figure*}

Towards the 94 target class I methanol maser sites we detect nine sites exhibiting 86.6- and 86.9-GHz emission. Spectra of each of these detections are presented in Fig.~\ref{fig:86_spect} and their properties are given in Table~\ref{tab:3mm_meth}. Of the nine detections, four appear to have typically thermal spectral profiles (G\,327.29$-$0.58, G\,351.77$-$0.54, G\,34.26+0.15 and W51E1), three have some indications of narrow, maser-like emission (G\,339.88$-$1.26, G\,344.23$-$0.57 and G\,351.42+0.65) and two appear to be maser emission (G\,345.01+1.79 and G\,29.96$-$0.02). As indicated in Table~\ref{tab:3mm_meth} five of these are new detections in these transitions, including one of the maser sources and two of the possible maser candidates. 

Five of the nine detections also show emission (or potential emission) in one or more of the 37.7-, 38.3- or 38.5-GHz transitions. These sources are discussed further in Section~\ref{sect:individual}.

\begin{table*}
\caption{Detections of methanol emission at 86.6- and 86.9-GHz. The first column gives the class I methanol maser source name and the next two groups of five columns give the minimum, maximum and peak velocity, the peak flux density and integrated flux density for the 86.6- and 86.9-GHz transitions, respectively. References to previously detected sources are: 1: \citet{Ellingsen03}; 2: \citet{Cragg01}; 3: \citet{Minier02} 
} 
  \begin{tabular}{lllllllllllllllllllllllllllll} \hline
 {Source name} & \multicolumn{5}{c}{86.6-GHz methanol masers} & \multicolumn{5}{c}{86.9-GHz methanol masers} & Refs\\
    {($l,b$)}& {V$_L$} & {V$_{H}$} & {V$_{pk}$} & {S$_{pk}$} & {S$_{I}$} & {V$_L$} & {V$_{H}$} & {V$_{pk}$} & {S$_{pk}$} & {S$_{I}$} \\	
    {($^{\circ}$ $^{\circ}$)} & \multicolumn{3}{c}{(\kmsns)} & (Jy) & (Jy \kmsns) & \multicolumn{3}{c}{(\kmsns)}& (Jy) & (Jy \kmsns)\\  \hline 
G\,327.29$-$0.58   &	$-$47.6 & $-$43.3 & $-$45.3 & 2.2 & 3.0 & $-$47.2 & $-$43.1 & $-$46.4 & 1.8 & 3.5\\
G\,339.88$-$1.26 & $-$38.4	& $-$33.6 & $-$38.2& 1.4 & 1.5 & $-$38.2 & $-$33.9 & $-$37.7 & 2.0 & 1.4	\\
G\,344.23$-$0.57 &   $-$25.1 & $-$16.4 & $-$21.8 & 2.8 & 4.2 & $-$23.3 & $-$16.8 & $-$23.2 & 2.0 & 2.7\\
G\,345.01+1.79 		&	$-$22.8 & $-$15.6 & $-$22.1 & 8.4 & 8.3 & $-$23.5 & $-$17.0 & $-$22.2 & 7.9 & 8.3 & 1,2 \\
G\,351.42+0.65 		&	$-$11.3 & $-$4.0 & $-$10.1 & 3.9 & 13.8 & $-$10.6 & $-$3.9 & $-$7.0 & 3.9 & 13.6 & 1,2\\
G\,351.77$-$0.54 	&	$-$7.8 & $-$1.0 & $-$4.2 & 4.4 & 14.3 & $-$7.9 & $-$1.1 & $-$7.1 & 4.2 & 15.2 & 2\\
G\,29.96$-$0.02		&	97.0 & 98.4 & 97.1 & 1.9 & 1.0 & 98.8 & 99.0 & 98.9 & 1.8 & 0.5\\
G\,34.26+0.15	&	57.0 & 61.4 & 58.6 & 2.7 & 6.7 & 56.1 & 61.2 & 59.5 & 2.6 & 7.8\\
W51E1			& 49.7 & 61.4 & 55.6 & 3.4 & 19.0 & 49.7 & 62.3 & 56.1 & 3.8 & 24.4	& 3\\ \hline
\end{tabular}\label{tab:3mm_meth}
\end{table*}


\subsection{Molecular and recombination line detections}

During our targeted methanol maser observations we were able to simultaneously observe a number of molecular and radio recombination lines. The parameters of these lines, derived from Gaussian fitting are presented in Table~\ref{tab:lines}. In the case where no emission is detected a 3-$\sigma$ detection limit is given. We only list H$^{13}$CN and H42$\alpha$ when the velocity range of the given source is included in the observations.

\begin{figure*}
	\epsfig{figure=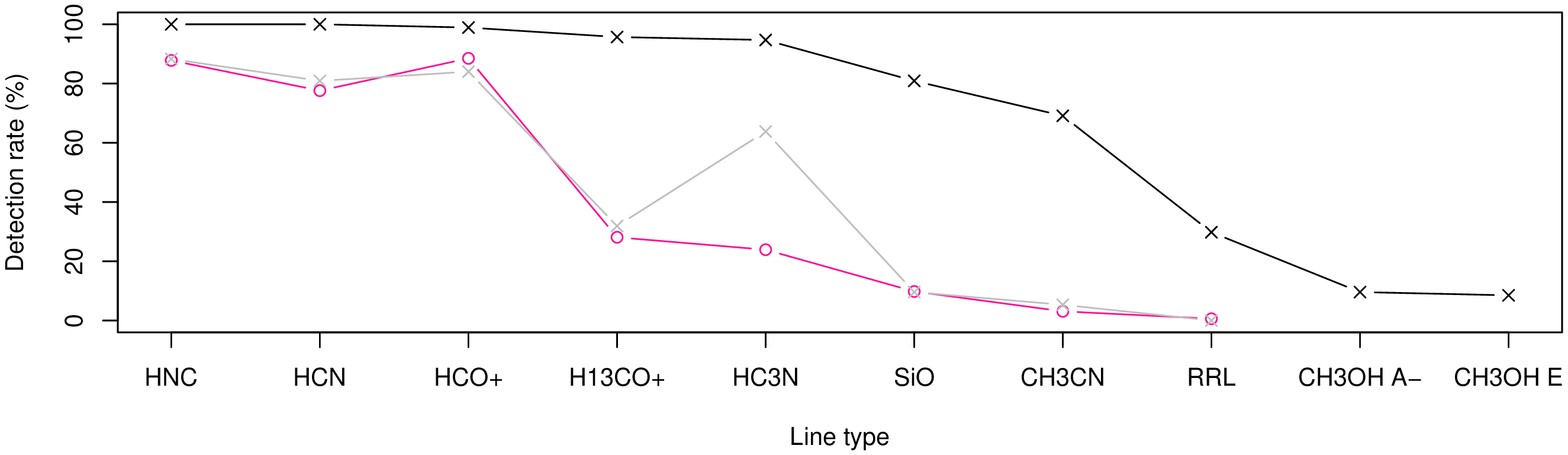,height=16cm,angle=270}
\caption{Detection rates (back crosses; corresponding to 100, 100, 98.9, 95.7, 94.7, 80.9, 69.1, 29.8, 9.6, 8.5\%) of the molecular and radio recombination lines observed. The detection rates of the MALT90 survey of 3246 dense clumps across the Galaxy are shown by pink open circles for comparison \citep{Rathborne16}. The MALT90 observations did not include the 88.9- (CH$_3$OH A$^-$) and 89.5-GHz (CH$_3$OH E) thermal methanol transitions so we cannot compare the detection rates for those two lines. The grey line shows what our detection rates would be at the MALT90 95 per cent completeness level \citep[T$_A^*$$>$0.4 K;][]{Rathborne16}.}
\label{fig:detection}
\end{figure*}

The detection rates of each of the lines which have a reasonably complete set of observations (i.e. the observing bandwidth accommodated the full complement of velocity ranges, so excludes H$^{13}$CN) are presented in Fig.~\ref{fig:detection}, showing that we detect HNC, HCN, HCO$^{+}$, H$^{13}$CO$^+$ and HC$_3$N towards 94.7 per cent of the methanol maser targets. SiO emission was detected towards 80.9 per cent of our target sources. CH$_3$CN was detected towards 69.1 per cent of the targets, although often not the full complement of hyperfine components, preventing us from deriving temperatures. Thermal methanol transitions 15$_3$ $\rightarrow$ 14$_4$A$^-$ and 8$_{-4}$ $\rightarrow$ 9$_{-3}$E are detected towards just 9 and 8 of the targets, respectively. One or more of the radio recombination lines were detected towards 29.8 per cent of the sample.	

\subsection{Comments on individual sources}\label{sect:individual}

In this section we draw attention to notable sources, associations, marginal detections and other details that are not able to be described fully in the source tables and spectra. \\

{\em S255 (G\,192.58$-$0.04).} \citet{Kurtz04} detected 44-GHz class I methanol maser emission towards this source using the VLA. \citet{Pratap08} conducted further single-dish observations of the 44-GHz emission in this source and used their derived position for targeted 36-GHz methanol maser emission. Their position was 49 arcsec offset from the \citet{Kurtz04} position, comparable to their HPBW, likely accounting for their non-detection in the 36-GHz transition. We detected 36-GHz methanol maser emission with a peak flux density of 12~Jy at the \citet{Kurtz04} 44-GHz methanol maser position. \\

{\em G\,300.97+1.15.} Although the candidate 37.7-GHz methanol maser emission we detect towards this source is relatively weak (2.1~Jy), its velocity is coincident with the 36-GHz methanol maser emission. \citet{Ellingsen11} targeted this site for 37.7-GHz methanol maser emission previously, achieving a 3-$\sigma$ detection limit of 5.7~Jy, significantly higher than the emission detected in the current observations. \\

{\em G\,318.95$-$0.20.} We detect a 37.7-GHz methanol maser towards this site, with a similar peak flux density to previous observations \citep{Ellingsen13,Ellingsen18}. While we failed to detect any emission from the 38.3- and 38.5-GHz transitions, \citet{Ellingsen18} report detections of these transitions with peak flux densities of 0.14 and 0.12~Jy, well below our 3-$\sigma$ detection limits.\\

{\em G\,323.74$-$0.26.} This site has been detected in the 37.7-GHz transition by \citet{Ellingsen11,Ellingsen13,Ellingsen18}, exhibiting some temporal variability between the observation epochs. The 2011 observations of \citet{Ellingsen18} found a peak flux density of 16.1~Jy, while the current observations detect emission of 32~Jy. The higher sensitivity observations of \citet{Ellingsen18} also allowed the detection of weak emission (0.28 and 0.22~Jy) in the 38.3- and 38.5-GHz transitions.\\

{\em G\,327.29$-$0.58.} Very marginal emission is detected towards this source at 37.7-, 38.3- and 38.5-GHz as well as more significant, thermal emission in the 86.6- and 86.9-GHz transitions. This is also one of the nine sources where the 88.9- and 89.5-GHz methanol transitions are detected. The fact that emission is detected in all the transitions, make the marginal emission in the 37.7-, 38.3- and 38.5-GHz lines believable. \\

{\em G\,335.79+0.17.} Both the 37.7- and 38.3-GHz masers we detect have been reported previously by  \citet{Ellingsen13} with peak flux densities of 13.8 and 7~Jy, respectively. We detect emission with the same peak velocity ($\sim$46.1~\kmsns) but with peak flux densities of 8.2 and 4.7~Jy, respectively. The target positions of the \citet{Ellingsen13} observations are within a couple of arcsec (so within the pointing uncertainty of the Mopra telescope) of the position we targeted so a pointing offset cannot account for the difference. \citet{Ellingsen13} commented that this source was unusual as the only source that has been found to have emission in the 38.3-GHz transition but not the 38.5-GHz transition. \\

{\em G\,339.88$-$1.26.} This site hosts the 37.7-GHz methanol maser with the highest flux density; 400~Jy in our observations. With their higher sensitivity observations, \citet{Ellingsen18} also detect emission in the 38.3- and 38.5-GHz transitions, with peak flux densities of 0.93 and 0.69~Jy.

In the 86.6-GHz transition we detect very marginal emission which is only reportable given the presence of slightly stronger 86.9-GHz emission, together with the previously reported tentative detection made at this site by \citet{Ellingsen03}. The current observations are not of sufficient sensitivity to be able to definitively rule out that some of the emission may be arising from a maser. Further, sensitive observations will be required to confidently infer its nature. We detect no emission from the 88.9- and 89.5-GHz methanol transition.   \\

{\em G\,343.12$-$0.06.} \citet{Voronkov06} conducted high spatial resolution observations of a number of class I methanol maser transitions towards this source, including the 84-GHz transition. \citet{Voronkov06} found that the 84-GHz methanol maser emission was spatially coincident with the 95-GHz transition, with some spots shown to be coincident with a molecular outflow.  \\

{\em G\,344.23$-$0.57.} \citet{Voronkov14} show that the 36- and 44-GHz class I methanol maser emission is distributed right out to the FWHM of the ATCA beam at 7mm (which is comparable to the Mopra beam), meaning that some of the 84-GHz components may lie beyond the half-power points of the smaller 3mm Mopra beam. The targeted position is close to the class II methanol maser location \citep{CasMMB11} and we detect a narrow feature at the 38.3-GHz methanol transition, but no emission in either the 37.7- or the 38.5-GHz transitions. The velocity of the detected emission is at $-$20.5~\kmsns, identical to the peak velocity of the 36-GHz transition. This velocity correspondence certainly gives credibility to the narrow emission. Even though it is unusual to see emission in the 38.3-GHz transition without corresponding 37.7 or 38.5-GHz detections, there are examples of other sources where 38.3-GHz emission is seen without accompanying 38.5-GHz emission (e.g. G\,335.79+0.17), and sources where the 37.7-GHz emission is significantly weaker than either the 38.3- and 38.5-GHz transitions (e.g. G\,351.42+0.65).

We also detect emission from the 86.6- and 86.9-GHz transitions, which is likely to be thermal but shows some hints of narrow emission in the 86.6-GHz line. Methanol emission in the 88.9- and 89.5-GHz transitions is also detected.  \\

{\em G\,345.01+1.79.} \citet{Ellingsen11} detected 37.7-, 38.3- and 38.5-GHz emission towards this source, with peak flux densities of 207, 9.4 and 5.0~Jy, respectively. Their high-resolution observations showed some variation in the peak flux density in the time between the two sets of observations, with peak flux densities of 181, 9.3 and 5.9~Jy for the respective transitions. We detected peak emission of 291, 44 and 21~Jy in our current observations, indicating significant variability between the 2011 observations of \citet{Ellingsen11} and our 2018 observations.

This source also hosts one of the few known examples of 86.6- and 86.9-GHz maser emission, first detected by \citet{Cragg01} with peak flux densities of 2.8 and 4.1~Jy at a velocity of $\sim$$-$21.7~\kmsns. Further 86.6-GHz observations by \citet{Ellingsen03} revealed two spectral features with flux densities of 16.4 and 10~Jy at velocities of $-$22.0 and $-$21.2~\kmsns. In our observations we find weak emission at $-$21.2~\kms of about 1.5~Jy and a main spectral feature at $-$22.1~\kms of $\sim$8~Jy at the two transitions. \\

{\em G\,351.42+0.65.} This source, also known as NGC6334F, has been observed at 37.3-, 38.3- and 38.5-GHz by \citet{Ellingsen11} and \citet{Ellingsen18} previously. Comparison of their peak flux densities with the current observations reveal some significant temporal variations in each of the 37.7-, 38.3- and 38.5-GHz transitions (with flux densities of 70, 39 and 107~Jy for the 37.7-GH transition; 174, 126 and 117 for the 38.3-GHz transition; and 150, 151 and 116 for the 38.5-GHz transition in the \citet{Ellingsen11}, \citet{Ellingsen18} and current observations, respectively).\\

{\em G\,9.62+0.19.} This is the site of the highest peak flux density 6.7-GHz methanol maser ever detected \citep[e.g.][]{Green12,Green17}. At 36- and 84-GHz we detected almost identical emission in the two transitions, with peak flux densities of 3.8 and 3.4~Jy, respectively. At 37.7-GHz we detected emission with a peak flux density of 30~Jy, slightly higher than previous observations \citep{Ellingsen11,Ellingsen13,Ellingsen18}. Observations by \citet{Ellingsen18} revealed weak emission in the 38.3-GHz transition (0.22~Jy), but no emission in the 38.5-GHz transition.  \\

{\em G\,10.6$-$0.4.} We detect very weak emission in the 38.5-GHz transition (peak flux density of 1.3~Jy) without detectable accompanying emission in either the 37.7- or 38.3-GHz transitions. The weak 38.5-GHz feature does share the same velocity of the peak feature in both the 36- and 84-GHz emission, lending some credibility to its authenticity. There are two nearby 6.7-GHz methanol maser sites \citep[both of which may be associated with the W31 region: G\,10.627$-$0.384 and G\,10.629$-$0.333;][]{GreenMMB10} with velocity ranges that overlap with the 38.5-GHz detection so it is unclear where the emission is located. This site was observed by \citet{Ellingsen11} but the emission was too weak (their rms is 1.1 Jy at 38.4-GHz) to be detected. Further, more sensitive observations would be needed to confirm this detection.\\

{\em G\,29.96$-$0.02.} We detect narrow emission in both the 86.6- and 86.9-GHz methanol transitions, making it the forth example of a maser in these transitions. The emission we detect show slightly different peak velocities at the two transitions, but the 86.6-GHz spectrum shows weak emission at the velocity of the 86.9-GHz maser peak. The peak velocity of the associated 6.7-GHz methanol maser is at 96.0~\kms \citep{Breen15}, slightly blueshifted compared to the 86.6- and 86.9-GHz detections, which at a velocity of $\sim$98~\kms still falls well within the overall 6.7-GHz velocity range of 93.4~ to 106.4~\kmsns.

\section{Discussion}

\subsection{Are the detected 36- and 84-GHz sources masers?}\label{sect:masers}

Fig.~\ref{fig:84_spect} shows a number of different spectral profiles, ranging from narrow maser-like features (e.g. G\,328.21$-$0.59), to broad, thermal-like components (e.g. G\,327.39+0.20) and a combination of the two (e.g. G\,331.13$-$0.24). From our single-dish observations alone, we can not definitively confirm that all of the sources that we detect are masers (or a combination of maser and thermal) since the implied lower limits on brightness temperature do not exceed the expectations for kinetic temperatures in high-mass star formation regions, but we can make some arguments that they are likely to be based on previous observations. The bulk of the target list (71/94) has been taken from \citet{Voronkov14} sample of southern class I methanol masers at 36- and 44-GHz. These sources have all been scrutinized with interferometric observations and confirmed to host maser emission at both 36- and 44-GHz \citep[the rest of the sample are known to host maser emission at 44-GHz][]{Kurtz04}. 

It is difficult to make a direct comparison of the 36-GHz spectra presented in Fig.~\ref{fig:84_spect} with those in \citet{Voronkov14} since the latter has high-resolution source maps which break the distributed maser emission into components while our spectra have blended all of the components into one single spectrum. However, comparison of some of the more ``thermal-looking'' (i.e. broader and more Gaussian-like) sources in Fig.~\ref{fig:84_spect} with the corresponding spectra in \citet{Voronkov14} show that our single-dish observations have have blended distinct maser components that are present in the higher-resolution data and that, in these sources, our flux densities are generally higher (although not by a consistent percentage). This suggests that the current single-dish spectra have additional thermal contributions that are resolved out in the interferometric observations. 

\citet{Jordan17} compared the spectral profiles of 44-GHz class II methanol masers derived from the auto- and cross-correlation data from the same interferometric observations taken with the Australia Telescope Compact Array. They found that, for their 77 maser sites, there was a huge distribution in the difference between the flux density of the two spectra, ranging from very similar to sources where about $\sim$70 per cent of the flux density was resolved out in the cross-correlation data (they had baselines up to 1.5~\kmsns).

Since it is clear that all of the 36-GHz sources in Fig.~\ref{fig:84_spect} that have been observed with interferometry (the majority) host maser emission, even those that have spectral profiles reminiscent of thermal emission in the current observations, it follows that the very similar 84-GHz spectra also contain maser emission, even in the case that they too appear to have thermal-like spectral profiles. However, comparing the 36-GHz flux densities of our current observations with those of \citet{Voronkov14}, combined with the findings of \citet{Jordan17} for 44-GHz class I methanol masers, it is also likely that in a number of cases our single-dish spectra also have contributions from thermal emission. 

\subsection{Previous 84-GHz observations: detection rates and nature of the emission}

Previous targeted class I methanol maser searches by \citet{Kalenskii01} and \citet{RG18} (both with single dish telescopes) have resulted in 84-GHz detection rates of 94 and 74 per cent, respectively. The latter target list was exclusively made up of 44-GHz targets, and the former is likely to be, although the authors just state that they are class I maser targets. \citet{Kalenskii01} reported that the majority of their sources (34/48) were likely to be quasi-thermal rather than simply maser emission based on their broad line-widths, especially when compared to their 44-GHz methanol maser counterparts. The spectral resolution (100~\kmsns) accounts for the slightly lower detection rate in  \citet{RG18} and prohibits them from assessing spectral profiles or intensities. 

Unlike \citet{Kalenskii01} who compared their 84-GHz spectra to the 44-GHz (7$_0$ $\rightarrow$ 6$_1$ A$^+$) transition, we find the spectral profiles for our two transitions to be remarkably similar. The 44-GHz transition is not in the same transition family as the 36- and 84-GHz transitions so is much less likely to have a similar spectral profile. Interferometric observations of 84-GHz methanol sources are very limited, but the one source in our sample that has been observed at high spatial resolution \citep{Voronkov06} was found to harbor maser emission. For these, and the reasons outlined in Section~\ref{sect:masers} we believe our sources contain maser emission, however, further high spatial resolution observations would be required to definitely understand the nature of the detected 84-GHz emission.



\subsection{Comparison of the 36- and 84-GHz spectral profiles}

Spectra of both the 36- and 84-GHz methanol maser lines are presented in Fig.~\ref{fig:84_spect} for each target, showing remarkably similar spectral profiles in the majority of cases. These 36- and 84-GHz methanol lines are the result of consecutive transitions of the same ladder, the next of which 
(the 6$_{-1}$$-$5$_0$E transition at 133-GHz) has also been found to closely resemble the spectral profiles of 84-GHz emission \citep{Kalenskii01}. While it is somewhat expected for pairs of transitions to show similar spectral profiles, our own observations show that the 36- and 84-GHz pair are much more alike than other cases such as the 86.6- and 86.9-GHz transitions, which are the next in sequence in the series that produces the 38.3- and 38.5-GHz lines. In another example, \citet{McCarthy18} used their interferometric data to compare the positions and velocities of both 44- (7$_0$ $\rightarrow$ 6$_1$ A$^+$) and 95-GHz (8$_0$ $\rightarrow$ 7$_1$ A$^+$) methanol maser features. They found that 49 per cent of 95-GHz maser components had accompanying 44-GHz emission whereas we find few examples of 84-GHz emission devoid of a 36-GHz counterpart (although higher spatial resolution would be needed to confirm this). \citet{Kim2018} conducted simultaneous 44- and 95-GHz observations, detecting 44-GHz emission towards 83 sources and accompanying 95-GHz methanol maser emission towards 68 of those, also indicating a slightly less close relationship than for the 36- and 84-GHz transitions.

\begin{figure}
	\epsfig{figure=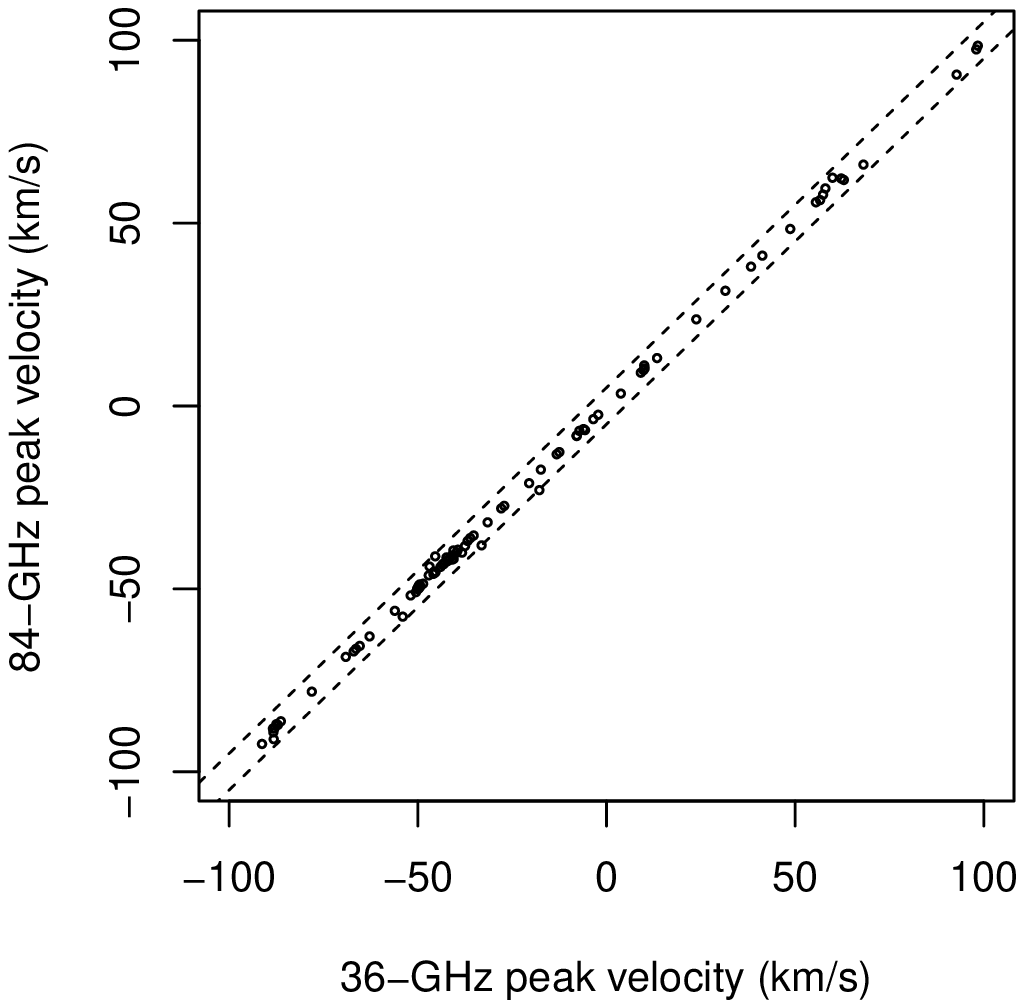,height=8cm,angle=270}
\caption{Peak velocity of the 36-GHz compared to 84-GHz methanol masers. The two dashed lines show a deviation of 5~\kms either side of unity.}
\label{fig:vel}
\end{figure}

Fig.~\ref{fig:vel} shows a comparison of the velocity of the peak 36-GHz and 84-GHz methanol emission, revealing a tight correlation between the peak velocities of the 92 sources detected at both frequencies. Only eight sources show 36- and 84-GHz peak velocity differences of more than 2~\kmsns, 75 show velocities within 1~\kms of each other and 67 of those are within 0.5~\kmsns. The largest peak velocity difference is 5.2~\kms in G\,345.42$-$0.95 which is a rare example of a source that shows no 84-GHz emission at the 36-GHz peak. Inspection of the \citet{Voronkov14} high spatial resolution data shows that the 36-GHz peak emission that we detected at $-$17.8~\kms lies well beyond the HPBW of the 3mm Mopra beam so further observations would be needed to rule out 84-GHz emission at that location. 

The mean and median velocity ranges are slightly higher for the 36-GHz sources compared to the 84-GHz sources - 10.1$\pm$0.8 and 9.2 compared with 7.7$\pm$0.5 and 7.1~\kmsns, respectively. The 36-GHz velocity ranges fall between 0.2 and 48.3~\kms and the 84-GHz sources between 0.6 and 34.8~\kmsns. The sources with the largest velocity ranges are G\,341.19$-$0.23 at 36-GHz and G\,328.81+0.63 at 84-GHz.


\begin{figure*}
	\epsfig{figure=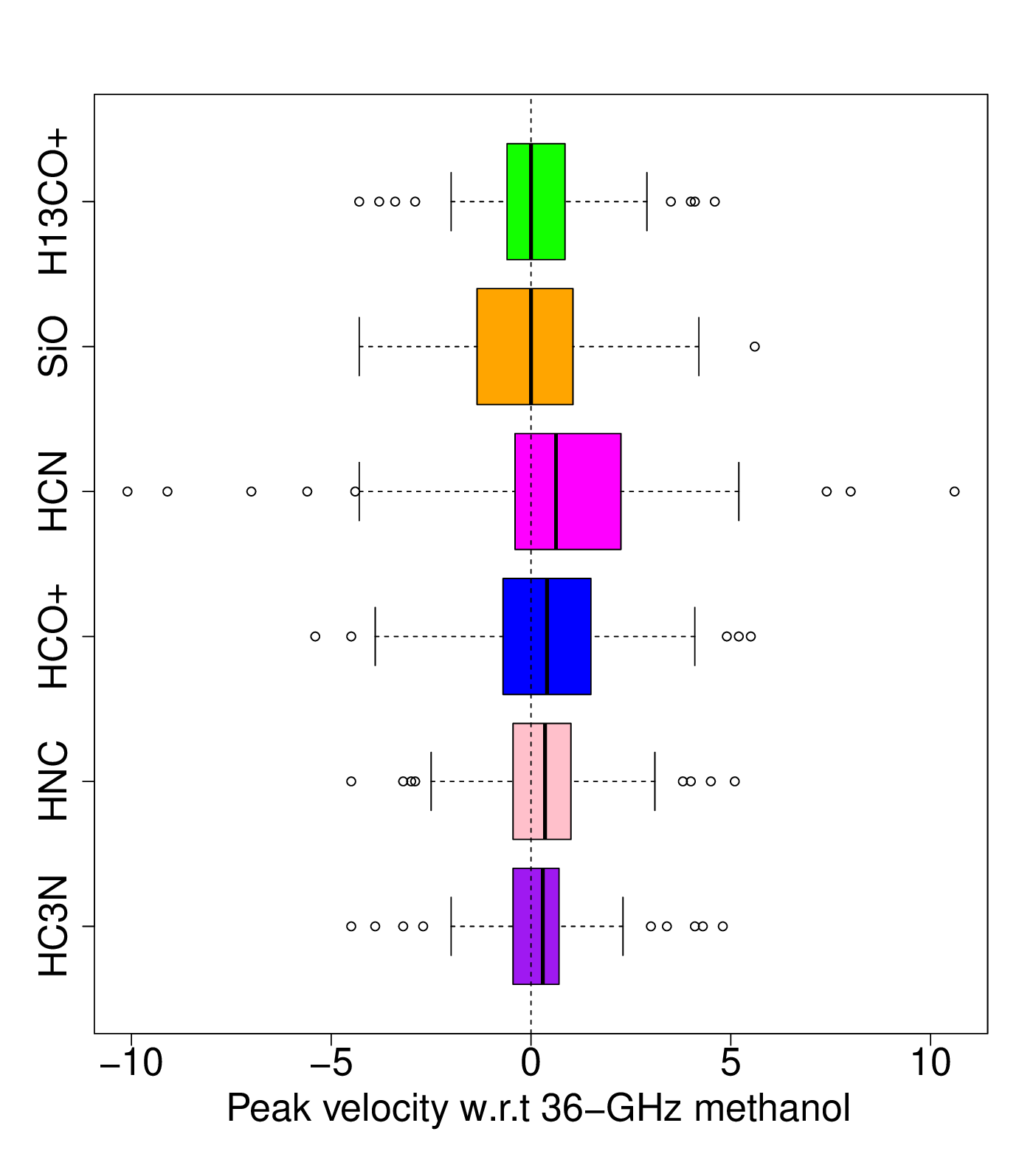,height=8cm,angle=270}
    \epsfig{figure=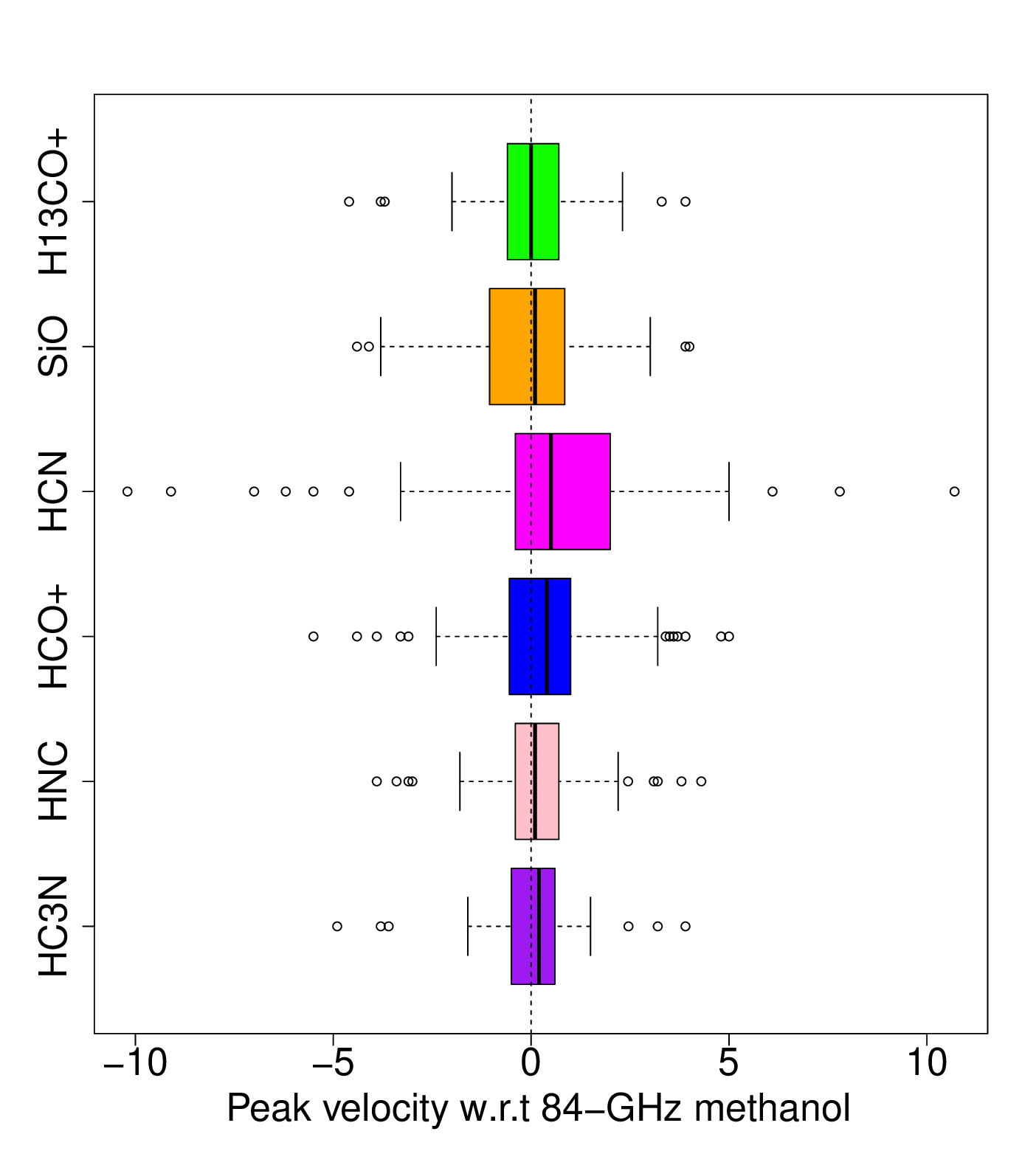,height=8cm,angle=270}
\caption{Box plots of the peak velocity of HC$_3$N, HNC, HCO$^+$, HCN, SiO and H$^{13}$CO$^+$ with respect to the peak velocity of the 36-GHz (left) and 84-GHz (right) emission. The vertical dashed line shows equal velocities. In each box plot, the
solid vertical black line represents the median of the data, the coloured
box represents the interquartile range (25th to the 75th percentile) and
the dashed horizontal lines (the ‘whiskers’) show the range from the 25th
percentile to the minimum value and the 75th percentile to the maximum
value, respectively. Values that fall more than 1.5 times the interquartile
range from either the 25th or 75th percentile are considered to be outliers
are represented by open circles.}
\label{fig:vels_thermal}
\end{figure*}

Fig.~\ref{fig:vels_thermal} shows box plots of the peak velocity of HC$_3$N, HNC, HCO$^+$, HCN, SiO and H$^{13}$CO$^+$ with respect to the peak velocity of the associated 36- and 84-GHz emission. In all cases the median velocity difference is close to zero, but it is clear that some molecules show better velocity agreement with the 36- and 84-GHz methanol masers than others. In most instances this is due to significant self-absorption in some of lines with high optical depths such as HCN, which is more often blueshifted with respect to the maser velocity, consistent with the detection of significant emission on the nearside of the source. From these comparisons it appears that the best maser velocity correspondence is with HNC, HC$_3$N and H$^{13}$CO$^+$ (the latter two are likely to be optically thin). 

In an analysis of 44-GHz class I methanol masers, \citet{Jordan17} found that class I methanol masers were better indicators of systemic velocities than class II methanol masers. Comparing the peak 44-GHz methanol maser velocity to CS (1--0) they found a mean velocity difference of 0.09$\pm$0.18~\kmsns, a median velocity difference of 0.04~\kmsns, and a standard deviation of 1.56~\kmsns. Table~\ref{tab:vel_diff} shows the mean, median and standard deviations of the velocity of the HC$_3$N, HNC and H$^{13}$CO$^+$ peak velocities with respect to the 36- and 84-GHz methanol maser peak velocities, showing that they are comparable to that found by \citet{Jordan17} when comparing the 44-GHz peak velocities to that of CS (1--0).

\begin{table*}
\caption{Mean (with standard error), median and standard deviations of the velocity of molecular lines with respect to the 36- and 84-GHz methanol maser velocities in units of \kmsns.} 
  \begin{tabular}{lllllllllllllllllllllllllllll} \hline
Line &  \multicolumn{3}{c}{36-GHz masers} & \multicolumn{3}{c}{84-GHz masers}   \\
   & mean	& median & s.d. & mean & median & s.d. \\ \hline
HC$_3$N & 0.22$\pm$0.16 & 0.29 & 1.46 & 0.05$\pm$0.12 & 0.2 & 1.17\\
HNC   & 0.30$\pm$0.16 & 0.35 & 1.54 & 0.13$\pm$0.14 & 0.1 & 1.32\\
H$^{13}$CO$^+$	& 0.10$\pm$0.16 & 0 & 1.51 & $-$0.04$\pm$0.13 & 0 & 1.24		\\ \hline
\end{tabular}\label{tab:vel_diff}
\end{table*}

\subsection{Comparison between the 36- and 84-GHz flux densities}

\begin{figure*}
	\epsfig{figure=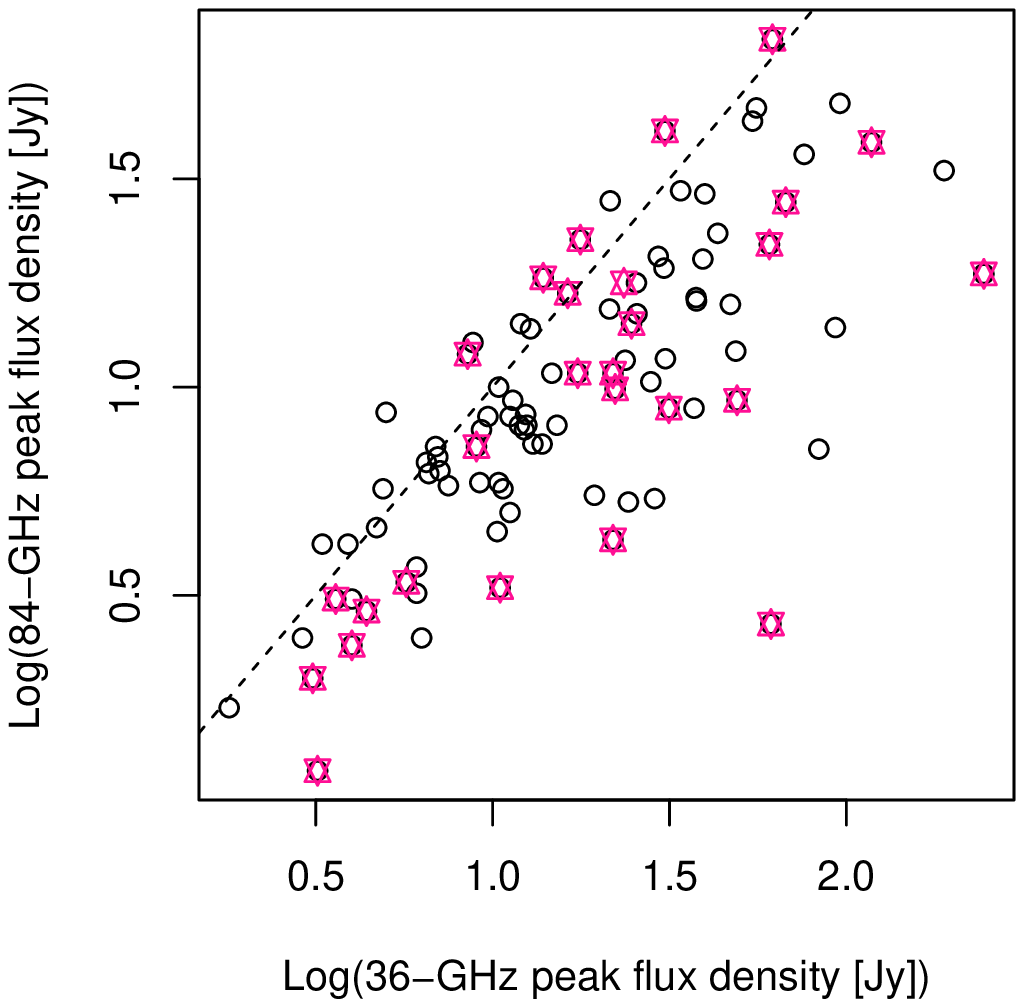,height=8cm,angle=270}
    	\epsfig{figure=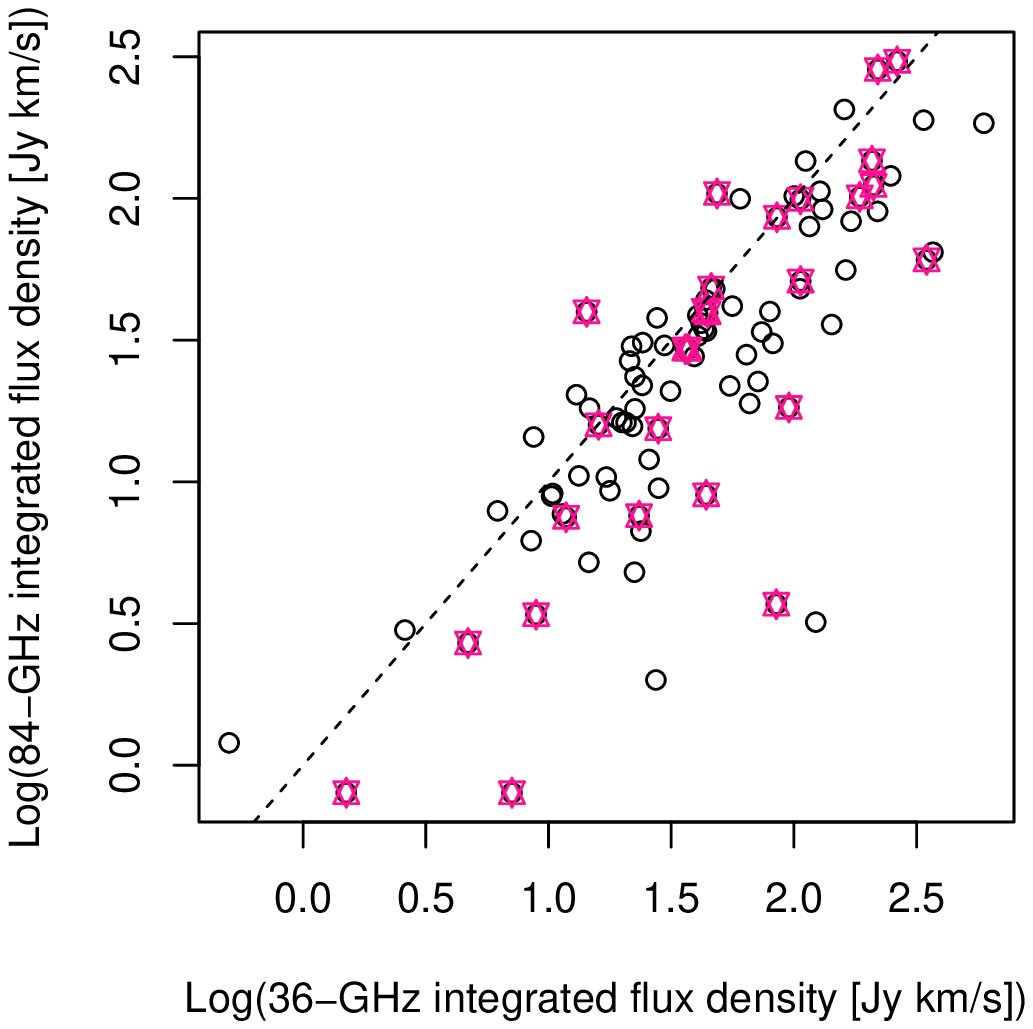,height=8cm,angle=270}
\caption{Log-log plots of the peak (left) and integrated (right) 36- versus 84-GHz methanol maser flux density. The dashed lines show x=y. Pink stars distinguish those sources that also exhibit radio recombination lines (28/92).}
\label{fig:flux}
\end{figure*}

The 92 36-GHz methanol maser detections range in peak flux density from 1.8 to 245~Jy (mean of 28.1, median 14.3~Jy) and the 93 84-GHz detections range in peak flux density from 1.2 to 68~Jy (mean of 13.0 and median of 8.7~Jy). The strongest 36-GHz maser is G\,335.59$-$0.29 and the strongest 84-GHz source is G\,327.29$-$0.58. A comparison between the 36- and 84-GHz peak and integrated flux densities are shown in Fig.~\ref{fig:flux}. The correlation coefficient between the peak and integrated flux densities of the two transitions are 0.52 and 0.71, indicating moderate and strong positive correlations. The fact that the integrated flux densities are more tightly correlated than the respective peak flux densities is reflected in Fig.~\ref{fig:flux} and indicates that the integrated flux densities are more robust to more extreme differences that might be seen in only a single velocity feature. The mean and median 36- to 84-GHz peak flux density ratio are 2.4 and 1.6, respectively. The 27 36-GHz methanol masers with peak flux densities that surpass 30~Jy have relatively higher 36- to 84-GHz peak flux density ratios with a mean value of 4.1 and a median of 2.4.

\begin{figure}
	\epsfig{figure=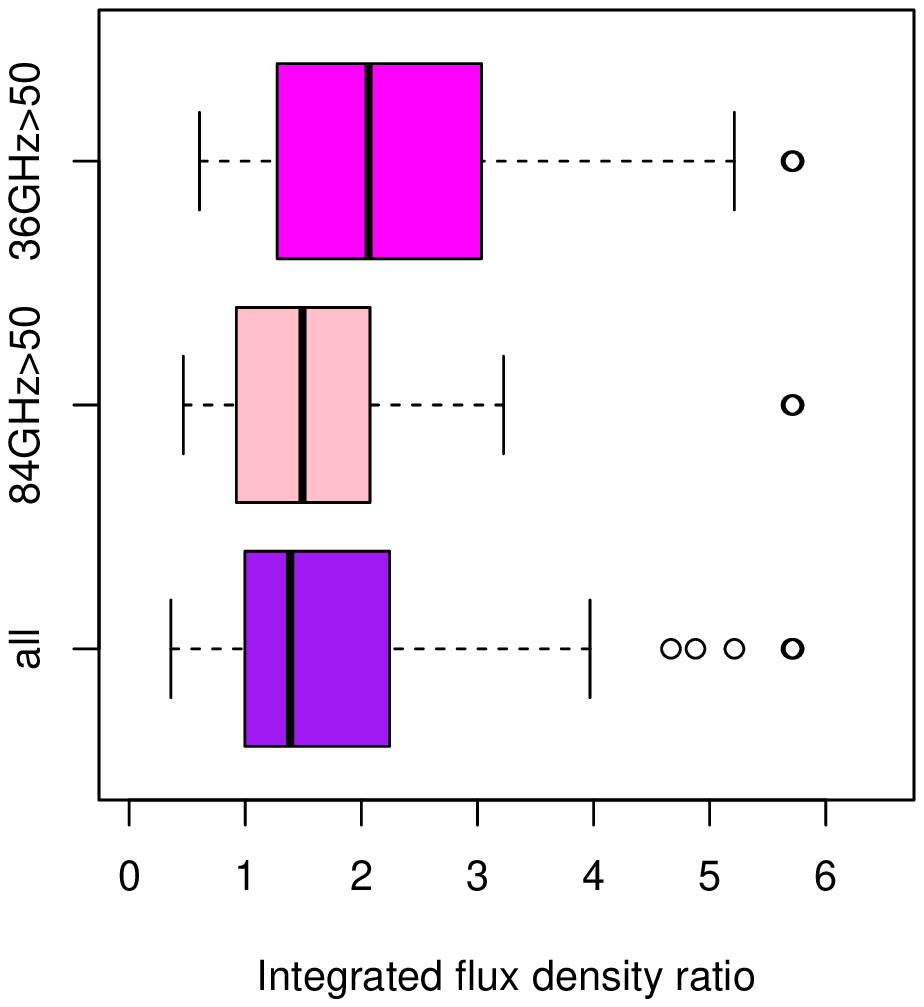,height=8cm,angle=270}
\caption{Box plots showing the ratio of 36- to 84-GHz integrated flux density for the full sample of 92 sources with detections at both frequencies (purple), the 24 sources with 84-GHz integrated flux densities greater than 50 Jy \kms (pink), and the 36 sources with 36-GHz integrated flux densities greater than 50 Jy \kms (magenta). Note that there are four sources with ratios greater than 6 (8.9, 13.7, 22.9 and 38.4), the 2 highest of which have 36-GHz integrated flux densities greater than 50 Jy \kmsns. See Fig.~\ref{fig:vels_thermal} for a general explanation of box plots.}
\label{fig:int_ratio}
\end{figure}

The average 84-GHz integrated flux density is 46.8 Jy~\kms and the median is 29.4 Jy~\kms compared with the 36-GHz sample which has an average of 77.0~\kms Jy and median of 41.3 Jy~\kmsns. The average 36- to 84-GHz integrated flux density ratio is 2.6 and the median is 1.4. The ratio of these lines is similar to that of the 84- to 133-GHz transitions, reported by \citep{Kalenskii01} to be 1.4. Fig.~\ref{fig:int_ratio} shows the distribution of integrated flux density ratios for the full sample, along with the ratio for both 36- and 84-GHz masers that have integrated flux densities of more than 50~Jy~\kmsns. For 36-GHz masers with integrated flux densities greater than 50~Jy~\kmsns, the mean and median ratio is 3.8 and 2.1, compared to  1.9 and 1.5 for the 84-GHz that have integrated flux densities greater than 50 Jy~\kmsns.

There are 16 cases where the 84-GHz peak flux density surpasses that of the 36-GHz methanol maser counterpart, 15 of which also have higher integrated intensities. There are an additional eight sources that have a larger 36-GHz peak flux density but have a higher 84-GHz integrated flux density. The sources with the largest 84- to 36-GHz integrated flux density ratio is G\,301.14-0.2. 

Although the 36- and 84-GHz methanol maser transitions have similar optimal conditions \citep[e.g.][]{Leurini16}, there are conditions that favor the 36- or 84-GHz transitions differently. For example, at high densities (10$^6$ - 10$^8$ cm$^{-3}$), 36- to 84-GHz ratios close to one might be expected but lower densities (10$^3$ - 10$^6$ cm$^{-3}$) can favor the 36-GHz transition, resulting in 36- to 84-GHz ratios of $\sim$2 \citet{McEwen14}. 

Fig.~\ref{fig:flux} also highlights the targets where radio recombination lines have also been detected. In previous studies, it has been suggested that other types of masers show a change in luminosity with evolution \citep[e.g.][]{Breen10} and if we consider the presence of detectable recombination line as an indication of a slightly more evolved site, Fig.~\ref{fig:flux} shows that there is no simple trend whereby the line ratio of the 36- to 84-GHz sources changes with evolution. Given that class I methanol masers trace shocks, they can be associated with multiple phases in the evolution of a young high-mass star \citep[such as outflows and expanding \ionhy regions;][]{Voronkov14} it is not surprising that there is no simple evolutionary trend. Further complicating the issue is the large Mopra beam which may lead to confusion between the multiple detected lines.

\subsection{Comparison of the class I methanol masers with thermal molecular lines}


Fig.~\ref{fig:detection} shows the detection rates of the thermal molecular and recombination lines we detect (excluding H$^{13}$CN since the frequency coverage excludes the velocities of most of the targets), highlighting the high association rates of most of the observed lines with our maser-associated star formation regions. The fact that HNC, HCN and HCO$^{+}$ were all detected towards 93 of the 94 target sources very strongly indicates the presence of dense gas at the locations of all of the target class I methanol masers. In the one case where HCO$^{+}$ is not reported in Table~\ref{tab:thermal} (towards G\,335.06$-$0.43), H$^{13}$CO$^+$ was, and there are hints of some narrow emission at the right velocity of the HCO$^{+}$ spectrum, indicating that it is significantly self-absorbed.

Also present in Fig.~\ref{fig:detection} are the MALT90 detection rates of eight of our lines toward a sample of 3246 high-mass clumps \citep{Rathborne16}. While MALT90 has similarly high detection rates of HNC, HCN and HCO$^+$, there is a significant drop off in the detection rates of H$^{13}$CO$^{+}$, HC$_3$N, SiO, CH$_3$CN and radio recombination line emission. This is mostly explained by the higher sensitivity of the current observations, as can be seen by the excellent agreement between the MALT90 detection rates and our detection rates adjusted to the MALT90 95 per cent completeness level, shown by the grey crosses in Fig.~\ref{fig:detection} \citep[T$_A^*$$>$0.4 K;][]{Rathborne16}. The notable exception is HC$_3$N, which has much higher detection rates towards the class I methanol maser selected sample. 
We have compared MALT90 detection rates as a function of source temperatures \citep[from][]{Guzman15} and also their MALT90-defined evolutionary categories, which indicate that, in order to account for the high HC$_3$N detection rates, the class I methanol maser targets are (in general) likely to be both protostellar and warm. 

\begin{figure*}
	\epsfig{figure=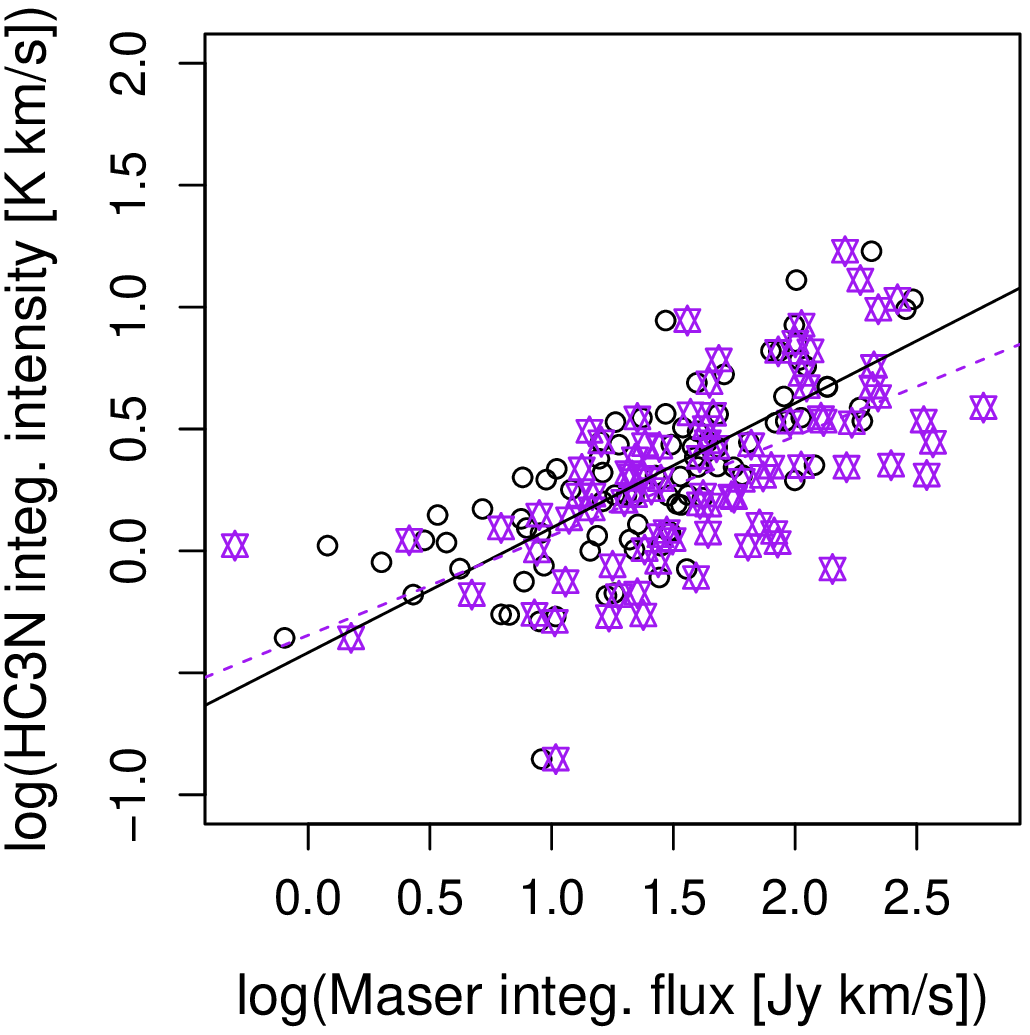,height=5.5cm,angle=270}
    \epsfig{figure=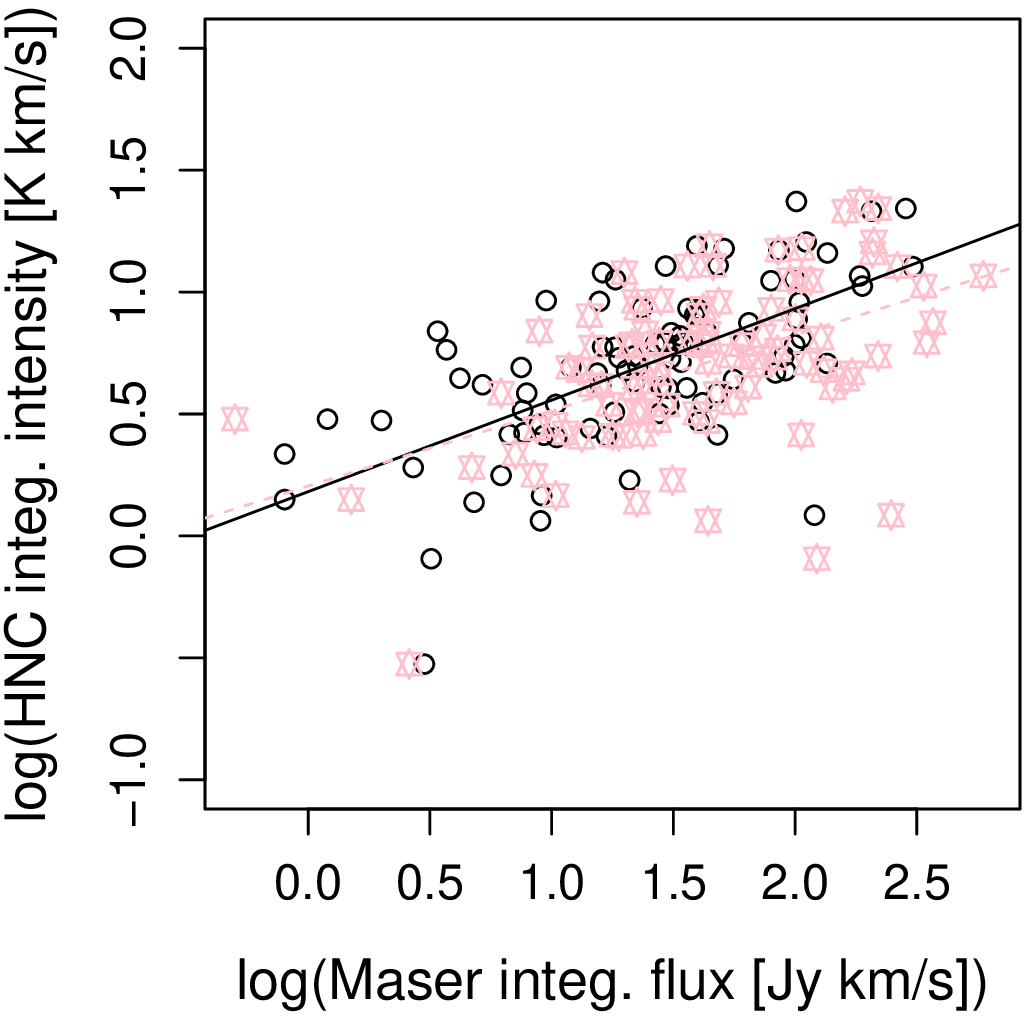,height=5.5cm,angle=270}
    \epsfig{figure=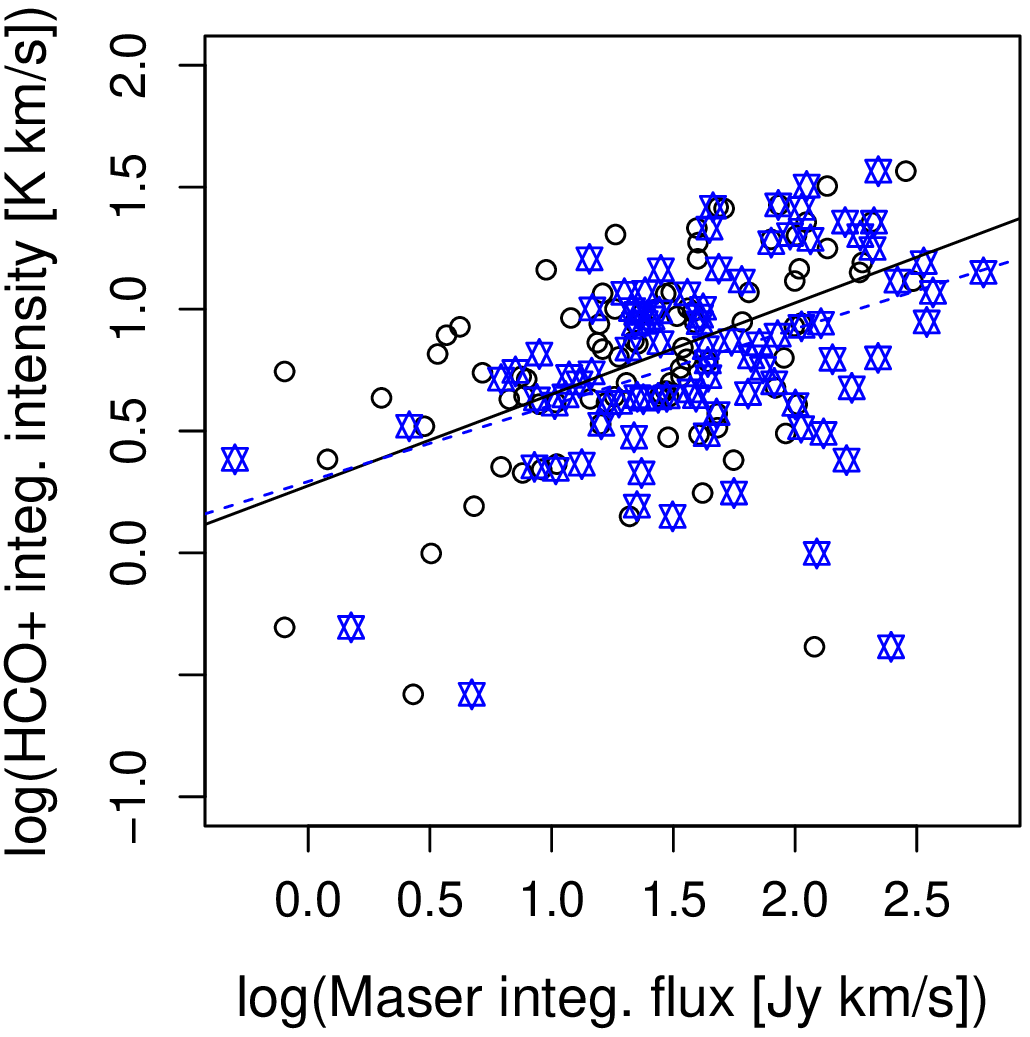,height=5.5cm,angle=270}   
\epsfig{figure=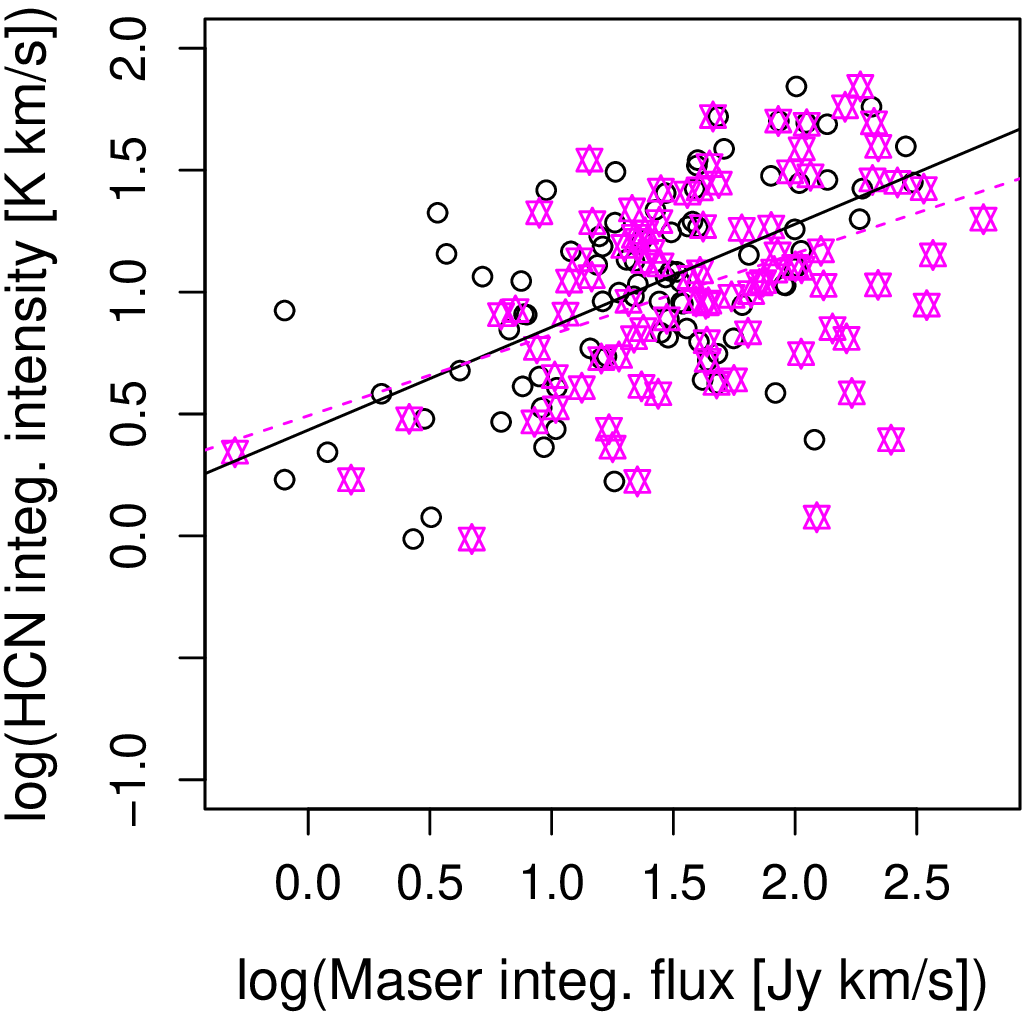,height=5.5cm,angle=270}
    \epsfig{figure=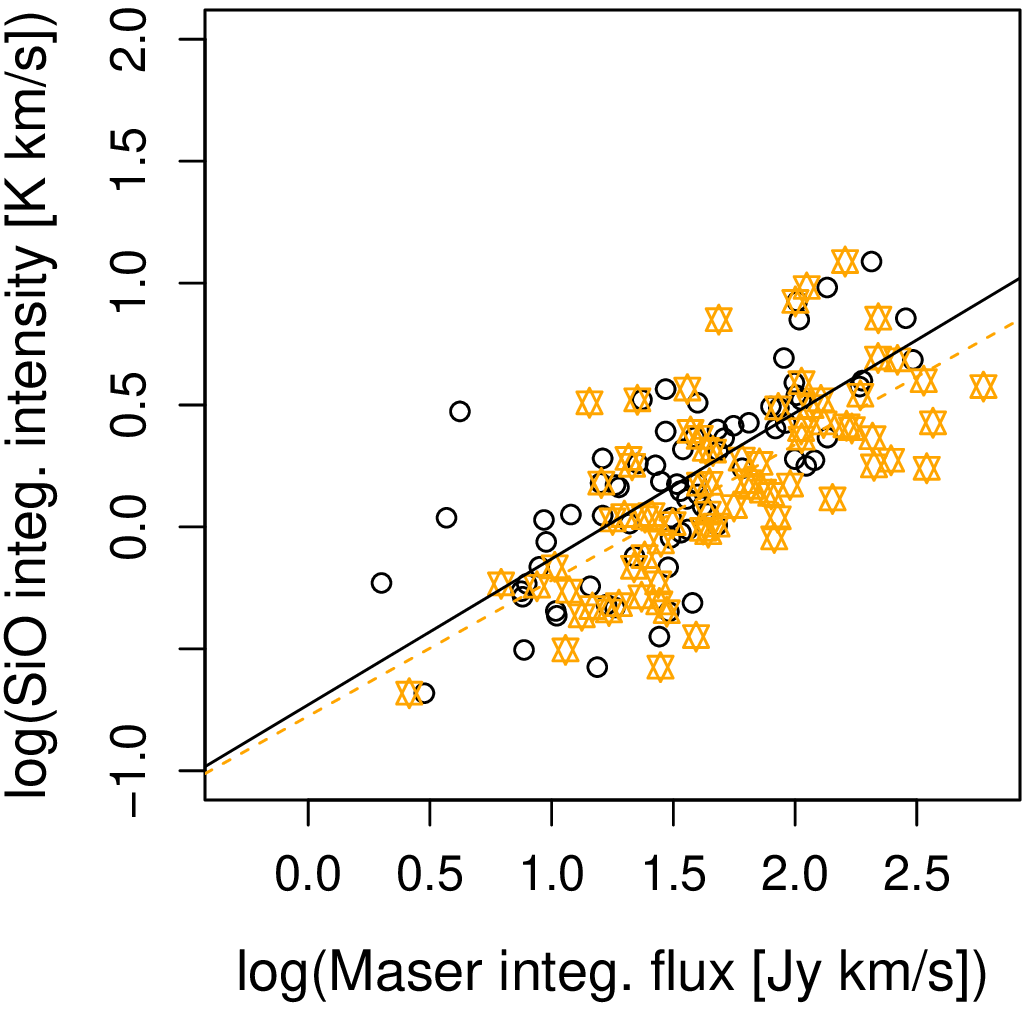,height=5.5cm,angle=270}
    \epsfig{figure=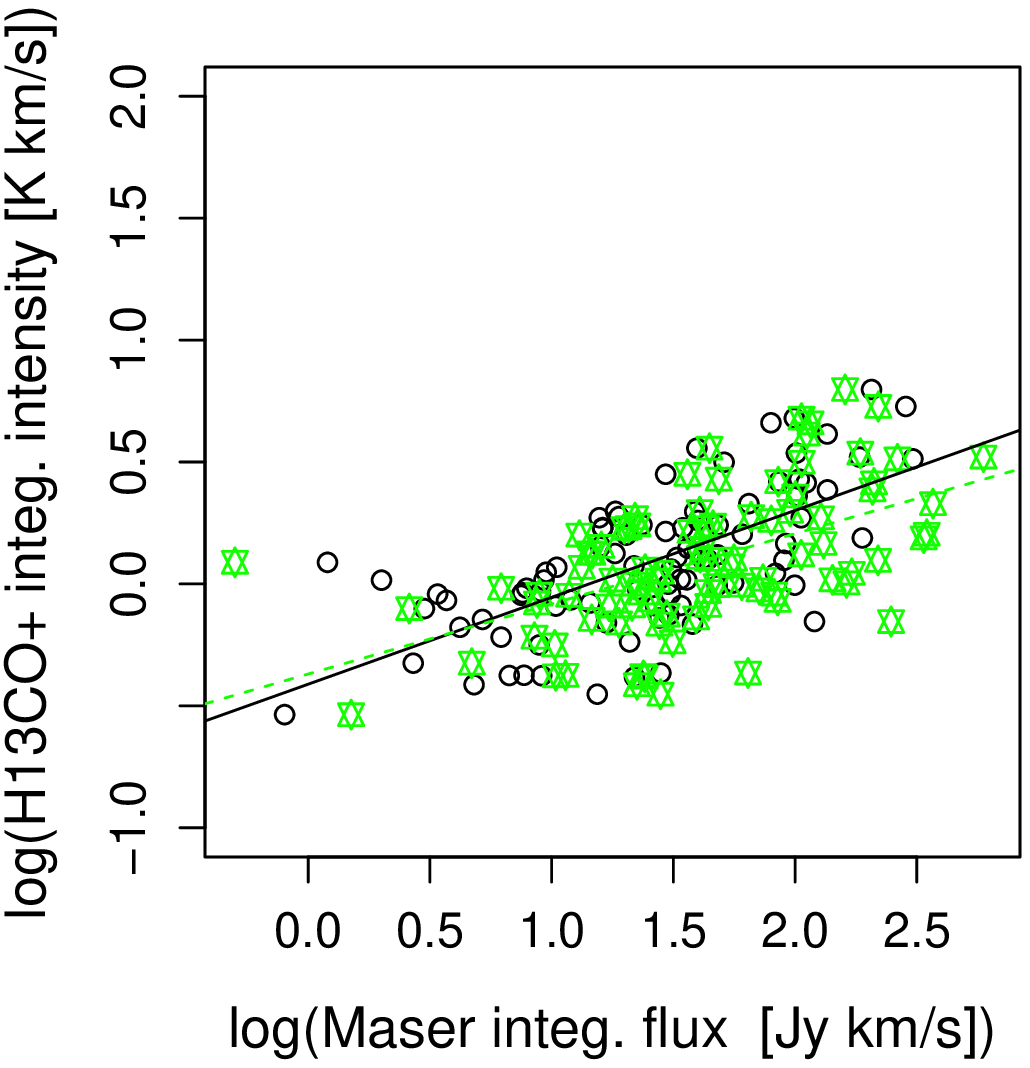,height=5.5cm,angle=270}
\caption{36- (coloured stars) and 84-GHz (black circles) methanol maser integrated flux density versus the integrated HC$_3$N, HNC, HCO$^+$, HCN, SiO and H$^{13}$CO$^+$ intensities. The lines of best fit is given in each case by a black line for the 84-GHz maser line and a matching coloured dashed line for the 36-GHz methanol masers. The Pearson correlation coefficients between each of the thermal and the 36-GHz transition are 0.61, 0.50, 0.42, 0.45, 0.69, 0.54 and 0.74, 0.63, 0.52, 0.58, 0.73, 0.66 for the 84-GHz transition.}
\label{fig:maser_integrated}
\end{figure*}

Fig.~\ref{fig:maser_integrated} shows the 36- and 84-GHz methanol maser integrated flux density plotted against the integrated HC$_3$N, HNC, HCO$^+$, HNC, SiO and H$^{13}$CO$^+$ intensities. The correlation coefficients mostly indicate moderately correlated positive relationships, suggesting that the 36- and 84-GHz sources with higher integrated flux densities are generally associated with molecular lines with higher integrated intensities. The slopes of the fitted linear relationship in each case are similar (0.41$\pm$0.06, 0.31$\pm$0.06, 0.31$\pm$0.06, 0.33$\pm$0.07, 0.56$\pm$0.07, 0.29$\pm$0.05 for the 36-GHz maser, and 0.51$\pm$0.05, 0.38$\pm$0.05, 0.37$\pm$0.07, 0.42$\pm$0.06, 0.60,$\pm$0.07 and 0.36$\pm$0.04 for the 84-GHz maser line), indicating that the integrated flux density of the maser lines scale with the overall quantity of gas. For optically thin gas, intensity scales linearly with the abundance of a given molecule, however, for the same temperature it has an exponential dependence on excitation energy. This, together with inadequate sensitivity, could account for the very low detection rate of the 88.9- (CH$_3$OH A$^-$) and 89.5-GHz (CH$_3$OH E) thermal methanol transitions. Furthermore, the spatial resolution of the current observations is insufficient to resolve the regions of gas where the masers originate, so determining the exact relationship would require higher-resolution follow up observations.



Comparison of the 44-GHz class I methanol integrated flux densities (derived from their auto-correlated data) with the integrated intensities of CS (1--0), SiO (1--0) and CH$_3$OH 1$_0$--0$_0$ A$^+$ by \citet{Jordan17} similarly found moderately correlated positive relationships (correlation coefficients of 0.41, 0.57 and 0.40, respectively) between the masers and the thermal line emission. They suggested that this could indicate that the more luminous 44-GHz methanol masers may be associated with the more massive high-mass star formation regions. Interestingly, \citet{Jordan17} find that the integrated intensity of the 44-GHz masers had a closer relationship with the integrated intensity of the SiO (1--0) emission than the other lines. In our targeted observations we find that the tightest linear relationships in Fig~\ref{fig:maser_integrated} are with HC$_3$N and SiO (2--1) which have Pearson correlation coefficients with the integrated 36-GHz methanol masers of 0.61 and 0.69, and with the integrated 84-GHz methanol masers of 0.74 and 0.73. This indicates that there is an even closer relationship between the SiO (2--1) and the 36- and 84-GHz integrated intensities than that of the 44-GHz methanol masers and SiO (1--0). The close intensity correlation between that of the collisionally excited class I methanol masers and SiO probably is a reflection of the fact that they are both tracers of shocked gas, often found in the vicinity of outflows \citep[e.g.][]{Garay02}. In the case of HC$_3$N, a hot core tracer, the positive linear correlation is likely suggesting that when there is a larger volume of hot and dense material (and so a higher HC$_3$N integrated intensity), there is also a larger volume of gas contributing to the maser emission. Alternatively, recent work by \citet{Taniguchi18} suggests that HC$_3$N can trace shocked gas so the tight correlation might also be reflecting a similar origin, as with the SiO emission.




Some molecular line ratios have been shown to change with the evolution of the associated high-mass star formation region \citep[e.g.][]{Hoq13,Sanhueza12}. \citet{Rathborne16} used the MALT90 sample to show that the HCO$^+$ to HNC and HCN to HNC integrated intensity ratios increased with evolutionary stage, agreeing with \citet{Hoq13} that this is probably a reflection of the fact that HNC is more abundant in less evolved clumps. \citet{Rathborne16} also found a similar trend in the HCO$^+$ to H$^{13}$CO$^+$ and HNC to HN$^{13}$C integrated intensity ratios and suggested that this is likely because there is either less self-absorption or a decrease in optical depth with clump evolution. The median values of our HCO$^+$ to HNC, HCN to HNC and HCO$^+$ to H$^{13}$CO$^+$ integrated line intensities are 1.3, 2.1 and 5.5, respectively. In their fig. 20, \citet{Rathborne16} show the median values for these line ratios in the categories of quiescent, protostellar and \ionhy regions (since they were looking for evolutionary trends). Our median line ratios are similar to those found in the quiescent, \ionhy regions and protostellar for the HCO$^+$ to HNC, HCN to HNC and HCO$^+$ to H$^{13}$CO$^+$ line ratios, respectively. This apparent discrepancy is probably a reflection of our smaller sample and the fact that median values are not a robust indicator of a distribution.

\begin{figure}
	\epsfig{figure=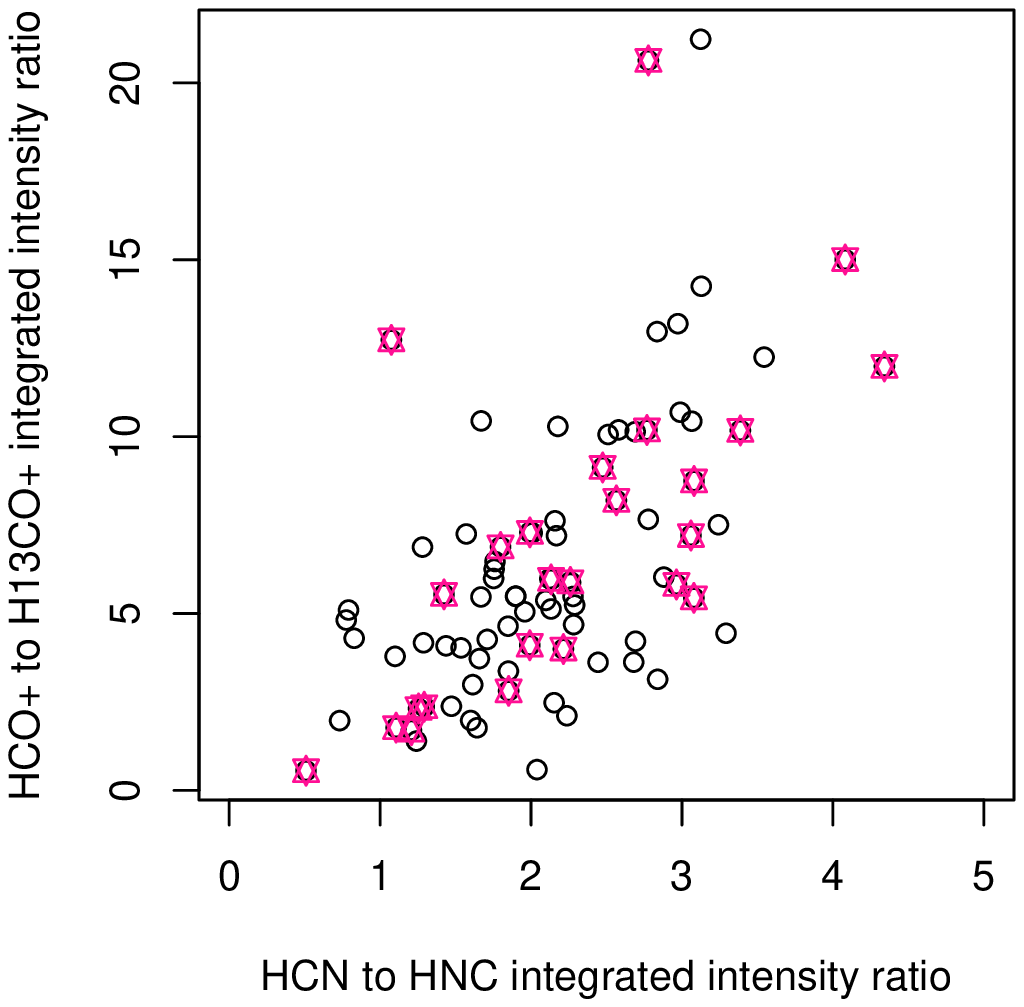,height=8cm,angle=270}
\caption{HCN to HNC versus HCO$^+$ to H$^{13}$CO$^+$ integrated intensity ratios (black circles). Sources where radio recombination lines are also detected are plotted by pink stars. Note that the x-axis has been truncated, excluding two sources with more extreme HCN to HNC ratios.}
\label{fig:mol_ratio}
\end{figure}

An investigation of the 36- and 84-GHz methanol maser properties with the ratios of HCO$^+$ to HNC, HCN to HNC and HCO$^+$ to H$^{13}$CO$^+$ revealed no obvious trends. Even though both our data and that from MALT90 suffer from confusion (given the large Mopra beam) it is possible that our much smaller sample prevents us from revealing a statistical change. Fig.~\ref{fig:mol_ratio} shows the HCN to HNC ratio plotted against the HCO$^+$ to H$^{13}$CO$^+$ ratio, revealing a weak positive correlation between the data points. Those targets where radio recombination lines are also detected are scattered throughout Fig.~\ref{fig:mol_ratio}, suggesting that the line ratios from our sample are not a good indication of evolution.

Aside from the common molecular lines which were also observed by MALT90, we also observed two thermal methanol transitions at 89.5~GHz (CH$_3$OH E) and 88.9~GHz (CH$_3$OH A$^-$). Their detection rates were relatively low (as can be seen in Fig.~\ref{fig:detection}) with both transitions detected towards eight sources and a further source detected in just the 88.9~GHz transition, but with a peak T$_A^*$ of just 0.04~K, below the 3-$\sigma$ detection limit of the 89.5~GHz transition. The upper energy level is 171~K for the 89.5~GHz transition and 328~K for the 88.9~GHz transition. The ratio of the 89.5~ to 88.9~GHz line integrated intensity falls between 0.76 to 1.3 for seven of the eight sources and is 7.1 for G\,327.29$-$0.58, the final source with detections in both transitions. For optically thin methanol gas in LTE these ratios imply rotational temperatures of $\sim$200~K for most of the sources, and $\sim$70~K for G\,327.29$-$0.58. Of the nine sources detected at either transition, five are associated with radio recombination lines. 

\subsection{Comparison of the class I methanol masers with recombination lines}

Compared to the MALT90 sample of dense clumps, we detected a larger fraction of sources associated with radio recombination lines (29.8 compared with 0.6 per cent) towards our sample of class I methanol masers. This difference can be fully explained by the fact that our data is not only more sensitive to H41$\alpha$ emission (the recombination line included in the MALT90 observations) but that we also observe a number of 7mm recombination lines (see Table~\ref{tab:lines}). In fact, if we consider only the H41$\alpha$ line and restrict our detection limit to the MALT90 95~per cent completeness level for peak T$_A^*$, as shown in Fig.~\ref{fig:detection}, the detection limits are consistent, indicating that there is no obvious bias towards the class I methanol maser targets being more evolved than a large fraction of the MALT90 sample.

Of the 28 sites of radio recombination line emission that we detect, many are associated with previously identified hyper and ultracompact \ionhy regions \citep[e.g.][]{Murphy10,Sewilo04,MH2003}. \citet{Murphy10} summerised the 
quantitative criteria for the discrimination between hyper and ultracompact \ionhy regions from the literature. They suggest that the consensus is that recombination line FWHM linewidths less than 40~\kms are a good indicator of ultracompact \ionhy regions, while FWHM linewidths greater than 40~\kms may suggest the presence of an hypercompact \ionhy region. Using this criterion, only three (G\,301.14-0.23, G\,5.89$-$0.39 and G\,34.26+0.15) of the 28 recombination lines are associated with hypercompact \ionhy regions, and the bulk of the detections (25/28) are associated with ultracompact \ionhy regions.


\subsection{37.7-, 38.3- and 38.5-GHz sources}


In the 37.7-, 38.3- and 38.5-GHz class II methanol maser transitions, we detected seven, three and two known sources, respectively \citep{Ellingsen11,Ellingsen13,Ellingsen18,Haschick89} plus five candidate lines towards four additional sources. As discussed in Section~\ref{sect:individual}, the velocities of the candidate sources often show corresponding emission in the 36-GHz transition, lending some credibility to their authenticity. \citet{Ellingsen18} compared the velocities of 37.7-GHz maser emission with that of the associated 6.7-GHz emission, finding that in the majority of cases, the 37.7-GHz emission was blueshifted with respect to the majority of the accompanying 6.7-GHz emission. Of the four targets where we detect maser candidates in the 37.7-, 38.3- and 38.5-GHz lines, all of them have velocities that are blueshifted with respect to the 6.7-GHz methanol maser peak \citep[reported in the MMB survey][]{GreenMMB10,CasMMB11,Green12}. 

\citet{Ellingsen13} looked at the temporal variability of 37.7-GHz methanol masers, comparing their 2012 data to data taken in 2009 by \citet{Ellingsen11}. Over the three year period they found that the largest changes in flux density was at the 40 per cent level. We have observed seven 37.7-GHz methanol masers that were included in \citet{Ellingsen13} observation and find that, in the 6 years between the observations, five of the sources have shown variations at the 15 - 41 per cent level (we see both increases and reductions at this level so it is not a calibration problem). The other two sources, G\,323.74$-$0.26 and  G\,351.42+0.65, have increased by 122 per cent and decreased by 52 per cent in that time.

We further find that 37.7-GHz sources with counterparts at 38.3-GHz and 38.5-GHz do not necessarily show similar level of variability in all of the detected transitions. In the case of G\,335.79+0.17, the 38.3-GHz transition has shown a reduction in peak flux density at approximately the same levels as for the 37.7-GHz transition (41 and 33 per cent, respectively). However, for G\,345.01+1.79 and G\,351.42+0.65, where we detect emission in all three of the lines we see an increase of 41, 219 and 198 per cent in G\,345.01+1.79 and an increase of 52 and decrease of 31 and 22 per cent in G\,351.42+0.65 for the 37.7-, 38.3- and 38.5-GHz transitions.


\subsection{86.6- and 86.9-GHz sources}

To date only three sites of 86.6- and 86.9-GHz maser emission have ever been reported \citep[towards G\,345.010+1.792, W3(OH) and W51-IRS1][]{Ellingsen03,Cragg01,Sutton01,Minier02}. Although the searches that uncovered these sources were not particularly extensive (\citet{Ellingsen03} targeted 18 sources, \citet{Cragg01} targeted 17 sources, \citet{Minier02} targeted 23 sources and \citet{Sutton01} only targeted W3(OH)) masers at 86- and 86.9-GHz are expected to be rare given that they are the next transitions up the ladder from the 38.3- and 38.5-GHz transitions (the 38-GHz transitions are 6$_2$ $\rightarrow$ 5$_3$A$^-$ and 6$_2$ $\rightarrow$ 5$_3$A$^+$ and the 86-GHz transitions are 7$_2$ $\rightarrow$ 6$_3$A$^-$ and 7$_2$ $\rightarrow$ 6$_3$A$^+$). Alongside the maser detections, these searches have uncovered a handful of thermal sources detected at either of these transitions \citep[Orion KL, NGC6334F, G351.77-0.54, W51E2 and NGC 7538-IRS1][]{Ellingsen03,Cragg01,Minier02}. \citet{Ellingsen03} also report the detection of marginal detections towards G\,323.740$-$0.263 and G\,339.884$-$1.259.

Our observations have uncovered nine detections of these transitions, four of which show no deviation from a typically thermal profile shape (G\,327.29$-$0.58, G\,351.77$-$0.54, G\,34.26+0.15, W51E1). A further three show possible narrow spectral features in one of the transitions (G\,339.88$-$1.26, G\,344.23$-$0.57, G\,351.42+0.65) but the noise levels in the current observations make it difficult to confirm if these are masers. We note that one of these (G\,351.42+0.65 or NGC6334F) has also been reported as a thermal source previously. The detection of G\,339.88$-$1.26 is very marginal at 86.6-GHz, but more convincing in the 86.9-GHz spectrum, allowing us to confirm the marginal detection from \citet{Ellingsen03}. 

We have detected two sources with convincing narrow features - G\,345.01+1.79 and G\,29.96$-$0.02, the latter of which is a new detection. Interestingly, G\,29.96$-$0.02 shows a different dominant spectral feature at the two frequencies, although a hint of emission is seen in the 86.6-GHz spectrum at the velocity of the 86.9-GHz detection. Higher signal-to-noise observations will be needed in order to show if the spectra are genuinely different.






Interestingly, of the nine sources detected in the 86.6- and 86.9-GHz methanol transitions, six were also detected in the 89.5 and 88.9 GHz lines. The three that have no associated 89.5 and 88.9 GHz emission are the most convincing maser detections - G\,339.88$-$1.26, G\,345.01+0.1.79 and G\,29.96$-$0.02. Five of the 86.6- and 86.9-GHz detections also show detections in one or more of the 37.7-, 38.3- and 38.5-GHz methanol transitions. Since the 86.6- and 86.9-GHz are the next transitions up the ladder from the 38.3- and 38.5-GHz transitions, we might expect that masers seen in the 38.3- and 38.5-GHz lines are good targets for 86.6- and 86.9-GHz masers.


\section{Summary}

We have surveyed a sample of 94 class I methanol maser sources \citep{Voronkov14,Kurtz04} for the little studied 84-GHz class I methanol maser transition. We also conducted near-simultaneous observations of the 36-GHz class I methanol maser transition to allow meaningful comparison of the two transitions and to derive line ratios. Alongside these observations, the flexibility of the Mopra spectrometer allowed us to concurrently search the sources for the rarer class II methanol maser transitions at 37.7-, 38.3-, 38.5-, 86.6- and 86.9-GHz as well as a number of thermal molecular and radio recombination lines. 

Towards the 94 class I methanol maser targets, we detected 84-GHz emission in 93 (all sources except Mol77) sources and accompanying 36-GHz emission towards 92 sources (all sources except Mol77 and G\,45.07+0.13). The spectral profiles of the two transitions are strikingly similar and we use this as the basis for an argument that our sources are likely to contain maser emission even in the case where the spectra are reminiscent of more typically thermal line profiles (since we know the 36-GHz transition shows maser emission from previous interferometric observations). The mean and median peak flux density 36- to 84-GHz ratio are 2.4 and 1.6, similar to the integrated flux density mean and median ratios of 2.6 and 1.4. We further find  that the stronger 36-GHz masers have higher 36- to 84-GHz ratios than the strong 84-GHz sources (as well as the full sample of sources). 

We detect one new source of 86.6- and 86.9-GHz methanol maser emission, adding to the small number of masers  that have been found in this transition. We detect a further known maser at 86.6- and 86.9-GHz, three sources that may contain narrow maser features and four sources that show no deviation from a thermal profile. In the 37.7-, 38.3- and 38.5-GHz transitions we detect emission (in one or more of the lines) from seven known sources and present four further maser candidates that require followup observations with higher sensitivity.

Comparison of the detection rates of thermal molecular lines toward our class I methanol masers with those found towards dense dust clumps across the Galaxy in the MALT90 shows almost identical rates in HNC, HCN, HCO$^+$, H$^{13}$CO$^+$, SiO, CH$_3$CN and the H41$\alpha$ recombination line (once the respective detection limits are accounted for). We, however, detect much higher rates of HC$_3$N, which we believe indicates that a larger proportion of the class I maser target list are warm protostellar sources compared to the MALT90 sample. 

We find a close correspondence between the peak velocity of the class I maser sources and the thermal line counterparts, in particular with HNC, HCO$^+$, H$^{13}$CO$^+$, supporting a result found previously for 44-GHz methanol masers, that class I methanol masers are generally excellent tracers of systemic velocities. 

There is a positive correlation between the 36- and 84-GHz integrated flux densities and integrated intensities of the detected thermal lines. Given the similarity of the slopes in each of the relationships, we suggest that this indicates that the maser integrated flux density is a reflection of the available quantity of molecular gas.

\section*{Acknowledgments}

The Mopra radio telescope is part of the Australia Telescope National Facility. Operations support was provided by the University of New South Wales, the University of Adelaide, The University of Sydney, The University of Newcastle, Nagoya University, NASA Goddard Space Flight Centre and Western Sydney University. This research has made use of: NASA's Astrophysics
Data System Abstract Service; and  the SIMBAD data base, operated at CDS, Strasbourg,
France. J.R.D. acknowledges the support of an Australian Research Council (ARC) DECRA Fellowship (project number DE170101086). S.P.E. acknowledges the support of ARC Discovery Project (project number DP180101061).

\appendix
\section{Thermal molecular and recombination line fits}
\onecolumn

\begin{longtable}{llllllrlrlllclllllllll}
\caption{Molecular and recombination line Gaussian fit parameters. The first column gives the source name, column 2 gives the line, column 3 gives the fitted peak antenna temperature, column 4 gives the peak velocity, column 5 gives the line width and column 6 gives the integrated intensity. In the case of CH$_3$CN, fits of each of the detected hyperfine components are presented on separate lines (velocities are all with respect to the rest frequency listed in Table~\ref{tab:lines}). Where a line is not detected in our observations, a 3-$\sigma$ detection limit is given in the place of the Peak T$_A^*$. Sources with an `t' in the place of a value of peak antenna temperature indicate that there is emission in the reference (as well as the source). In one case there is a `d' in the place of peak antenna temperature indicating that there is weak emission present but it is unable to be fit by a Gaussian.  } 
\label{tab:thermal}\\ \hline
\multicolumn{1}{l}{Source name} & Line & Peak T$_A^*$ & V$_{peak}$ & V$_{width}$ & Integrated \\  
          & & (K) & (\kmsns) & (\kmsns) & (K~\kmsns)\\ \hline 
S255$^2$	     	&	CH$_3$CN 	& 0.16(0.03) & 1.93(0.09) & 0.93(0.21) & 0.154\\ 
& &0.14(0.02) & 3.61(0.14) & 1.72(0.37) & 0.265\\
					 & 	HC$_3$N	 	& 0.77(0.01) & 8.64(0.03 &3.23(0.06) & 2.667\\
                     &  HNC 	 	& 1.86(0.01) & 8.65(0.01) & 3.14(0.03) & 6.219\\
                     & CH$_3$OH E 	& $<$0.14\\
\hline

\end{longtable}

\twocolumn

\end{document}